\documentclass[extra] {gji}

\pdfoutput=1

\usepackage{xcolor}
\usepackage[colorlinks=True,%
            urlcolor=blue,%
            linkcolor=blue,%
            citecolor= blue,%
            pdfauthor={O. Barrois},%
            pdftitle={},%
            pdftex]{hyperref}
            
\usepackage{timet}
\usepackage{graphicx}
\usepackage{amsmath,amsfonts,amssymb}
\usepackage{longtable}
\usepackage{caption}
\usepackage{bm}

 \usepackage{lineno}

\renewcommand{\cite}{\citet}

\title[QG convection-driven dynamos]
	{Quasi-geostrophic convection-driven dynamos in a thick spherical shell}
\author[Barrois et al.]
  {O. Barrois$^1$, T. Gastine$^2$, C.~C. Finlay$^1$ \\
  $^1$ Division of Geomagnetism, DTU Space, Technical University of Denmark, Lyngby DK-2800, Denmark.\\
  $^2$ Universit\'e Paris Cit\'e, Institut de Physique du Globe de Paris, CNRS, F-75005 Paris, France.}
\date{Received XXX; in original form XXX}
\pagerange{\pageref{firstpage}--\pageref{lastpage}}
\volume{XXX}
\pubyear{2021}

\begin{document}

\label{firstpage}

\maketitle

\begin{summary}

We present dynamos computed using a hybrid QG-3D numerical scheme in a thick spherical shell geometry.
Our model is based on a quasi-geostrophic convection code extended with a 3D treatment of heat transport and magnetic induction.
We find a collection of self-sustained, multipolar, weak field dynamos with magnetic energy one or two orders of magnitude lower than the kinetic energy. 
The poloidal magnetic energy is weak and, by construction, there is a lack of equatorially anti-symmetric components in the Buoyancy and Lorentz forces.
This leads to configurations where the velocity field is only weakly impacted by the magnetic field, similar to dynamos found in 3D simulations where zonal flows and the $\Omega$-effect dominate.
The time-dependence of these dynamos is characterised by quasi-periodic oscillations that we attribute to dynamo waves.
The QG-3D dynamos found so far are not Earth-like.
The inability of our setup to produce strong, dipole-dominated, magnetic fields likely points to a missing ingredient in our QG flows, and a related lack of helicity and $\alpha$-effect.
The models presented here may be more relevant for studying stellar dynamos where zonal flows are known to dominate.
This study was carried out at modest control parameters, however moving to lower Ekman numbers, when smaller values of both the magnetic and hydrodynamic Prandtl numbers can be of interest, our approach will be able to gain in efficiency by using relatively coarse grids for the 3D magnetic and temperature fields and a finer grid for the QG velocity field.

\end{summary}


\section{Introduction}
\label{sec:state_of_art}

The magnetic field of the Earth is produced and sustained in Earth's outer core by turbulent motions of the liquid metal.
These motions are widely believed to be driven by thermal and chemical convection and can be described by the equations of magnetohydrodynamics (hereafter MHD).
The rotation of the Earth strongly influences the dynamics of the outer core as is evident from the extremely small values of its Ekman number, that characterises the ratio between the viscous and the Coriolis forces, $Ek = \nu / \Omega d^2 \sim 10^{-15}$ -- where $\nu$ is the kinematic viscosity, $\Omega$ is the Earth's rotation rate and $d$ is the thickness of the outer core -- and its Rossby number, which characterises the ratio between the inertia and the Coriolis forces, $Ro = U / \Omega d \sim 10^{-6}$ -- where $U$ is a typical velocity of the fluid.
In addition to the inertia and the viscosity, which are dwarfed by the Coriolis force, the remaining forces in the system are the Buoyancy and the Lorentz force.
The fact that $Ek$ and $Ro$ are so small in the outer core has lead to suggestions that it may be in a Quasi-Geostrophic (hereafter QG) dynamical balance at the $0^{th}$ order, {\it i.e.} a balance between the pressure and the Coriolis force \citep{davidson2013scaling,calkins2018quasi}.
Other authors have argued that a Magnetostrophic balance could hold at leading order, {\it i.e.} an equilibrium between pressure, Coriolis and Lorentz forces \citep{roberts1965analysis,dormy2016strong}; the relevant balance may depend on the length-scale \citep{aurnou2017cross,schwaiger2019force} and part of the fluid volume \citep{schaeffer2017turbulent} considered, the QG-balance being dominant in most cases at large length-scales and outside of the tangent cylinder and of the boundary layers.

Numerical simulations of the primitive equations governing core dynamics have proven to be powerful tools for investigating dynamo mechanisms in a spherical shell geometry and their parameter dependencies \citep[{\it e.g.},][]{glatzmaier1995three, christensen2006scaling}.
Despite being restricted to a region of parameter space still remote from that of the Earth's core -- $Ek \lesssim 10^{-7}$ and $Re \lesssim 5\times 10^3$ to be compared with $Ek \sim 10^{-15}$ and $Re \sim 10^9$, where $Re = U d / \nu$ is the Reynolds number characterising the ratio between the inertia and the viscous force -- progress has gradually been made in moving towards the Earth-like regime and in understanding the generic mechanisms at work  \citep{schaeffer2017turbulent,sheyko2018scale,aubert2017spherical,schwaiger2019force}.
In particular, following a specific path in parameter space, holding some parameters fixed at Earth-like values and gradually moving others toward the desired values, \citet{aubert2019approaching, aubert2023state} has been able to approach the conditions of Earth's core, by focusing on large length-scales and employing hyperdiffusion.
Such 3D simulations are nonetheless a major computational undertaking, especially for the most extreme parameters.

The possibility that the Coriolis force dominates and that core motions are columnar has motivated the development of reduced QG models to study core dynamics.
These involve a 2D projection of the 3D MHD equations with the flow dynamics of the system being constrained to the equatorial plane \citep{busse1970thermal,cardin1994chaotic,gillet2006quasi}.
Despite some limitations, especially if the temperature is also treated as 2D and contributions from the thermal wind are not included \citep{gillet2006quasi}, this approach has proven effective at mimicking the behaviour of the full 3D system, at least in the purely hydrodynamic case \citep{aubert2003quasigeostrophic,guervilly2019turbulent,barrois2022comparison}.
In the past few years, some interesting extensions of the QG method, mainly aiming at better accounting for the dynamics near the outer boundary, have been proposed \citep{labbe2015magnetostrophic,maffei2017characterization,jackson2020plesio,gerick2020pressure}.

A small number of studies have tried to include the effect of the magnetic field within a QG framework.
QG-MHD models have been implemented in the context of mechanical forcing and only considering the largest scales of the Lorentz force \citep{schaeffer2006quasi} or within a kinematic dynamo framework with convection including the effects of a 3D thermal wind \citep{guervilly2010thesis}.
\citet{schaeffer2006quasi} argued that a combination of flow time-dependence and the $\beta$-effect -- due to the Coriolis force acting on fluid columns in spherical geometry -- is sufficient to produce a QG dynamo independent of Ekman pumping.
Both studies managed to obtain dynamos but found that the onset of the dynamo action as a function of the magnetic Reynolds number $Rm = U d / \lambda$, which characterises the ratio of magnetic diffusion over the convection overturn timescales -- where $\lambda$ is the magnetic diffusivity -- was about five to ten times higher than that expected for 3D dynamos \citep[{\it e.g.},][]{petitdemange2018systematic}, with critical values of $Rm_c \sim 500$ in the QG compared with $Rm_c \sim 50$ in the full 3D case.

Several studies have investigated eigenmodes in QG-MHD systems, including the effect of the Lorentz force, considering small perturbations about an imposed background magnetic field.
\citet{canet2014hydromagnetic} considered only the dynamics of axially invariant magnetic fields within a purely QG model, while more recently hybrid models have been developed, considering QG flow but a 3D magnetic field and projecting the 3D quantities on a QG basis \citep{gerick2021fast}.
Lately \citet{jackson2020plesio} have described a more complete 2D model QG-MHD based on quadratic magnetic quantities.

Using mean-field electrodynamics theory \citep{steenbeck1966berechnung,krause1980mean}, it is possible to characterise dynamo action by considering azimuthally-averaged effects \citep[see {\it e.g.}][]{schrinner2007mean}.
In this context, the terminology $\alpha$-effect refers to the mean electrodynamics effect of helical flow generating poloidal magnetic energy from toroidal magnetic energy (or toroidal magnetic energy from poloidal energy) while $\Omega$-effect refers to the production of toroidal from poloidal magnetic energy through an axisymmetric shear flow \citep{parker1955hydromagnetic,moffatt1978magnetic,hollerbach1996theory}.
Dynamo action in 3D convection-driven models of the geodynamo that produce strong dipolar fields is usually classified as being of $\alpha^2$ type \citep{olson1999numerical}, at least when considering field generation by convective motions outside the inner core tangent cylinder.
Inside the tangent cylinder the $\Omega$-effect can also play a role, especially in strongly-driven cases \citep{schaeffer2017turbulent}.
In contrast, when strong zonal winds dominate convection outside the tangent cylinder the dynamo mechanism is typically found to be of $\alpha\Omega$ type with the resulting poloidal magnetic fields being weak and multipolar \citep{schrinner2012dipole}.
It seems there is a trade-off between strong zonal winds and strong dipolar magnetic fields.

QG models can efficiently simulate the dynamics of strong zonal flows \citep{schaeffer2005quasigeostrophic, gastine2019pizza}, so they might be expected to be relevant for studying dynamos where the $\Omega$-effect is important, for instance in the context of stellar magnetic fields \citep[{\it e.g.},][]{grote2000hemispherical,goudard2008relations}, or gas giants \citep[{\it e.g.},][]{gastine2012dipolar}.  It is however less obvious whether or not QG dynamo models are relevant to terrestrial planets such as the Earth \citep[{\it e.g.},][]{aubert2013bottom}.
\citet{schaeffer2016can} have shown, within a kinematic dynamo framework, that adding magnetic pumping \citep[an additional source of helicity related to the action of the Lorentz force, see][]{sreenivasan_jones_2011} enables simple, observation-based, QG flows to generate dipole-dominated dynamos. The question of whether dynamically-consistent QG flow models, driven by convection and including feedback from the Lorentz force, can result in Earth-like dynamos is central to our study.

Our main objective here is to develop a hybrid QG-3D model based on QG convection in a thick spherical shell geometry \citep{gastine2019pizza}, incorporating a 3D temperature field and thermal wind effects \citep{guervilly2017multiple,barrois2022comparison}, treating the magnetic field and its time evolution through the magnetic induction equation in 3D, and exploring the type of dynamos that are possible in this configuration.
We compute the Lorentz force in 3D then $z$-average to obtain the impact on the QG flows.
Building on previous QG-dynamos studies \citep{guervilly2010thesis,schaeffer2006quasi,schaeffer2016can}, we present here an attempt to produce fully-resolved self-consistent convection-driven dynamos.

We describe our method and then the equations used in Section~\ref{sec:Method}, present our main results in Section~\ref{sec:Results} and we conclude with a brief discussion and summary in Section~\ref{sec:Conclusion}.  Tables with diagnostics and benchmarks of our method can be found in the Appendix section.

\section{Methodology}
\label{sec:Method}

\subsection{Hybrid QG-3D model  formulation}
\label{sec:QGH-mag_equations}

Our hybrid QG-3D model builds on earlier work by \citet{schaeffer2005quasigeostrophic,gillet2006quasi,guervilly2016subcritical} and \citet{gastine2019pizza}.
We adopt the QG model formulation and notations of \citet{barrois2022comparison}, and use the cylindrical coordinates system $\left( s, \phi, z \right)$ -- with unit vectors $\left( {\bm e}_s, {\bm e}_\phi, {\bm e}_z \right)$ -- in a spherical shell between the inner and outer radii, $s_i$ and $s_o$ respectively, that rotates about the $z$-axis at a constant angular velocity $\Omega$.
We take $\eta = s_i / s_o = 0.35$ suitable for a thick shell such as the Earth's outer core.
We solve the dimensionless equations of our problem under the Boussinesq approximation for the velocity field ${\bm u}$, the magnetic field ${\bf B}$ and the temperature field $T_\text{3D} \equiv T_\text{3D}^\text{cond} + \vartheta_\text{3D}$.
The last two fields are fully treated in 3D using the spherical coordinates system system $\left( r, \theta, \phi_\text{3D} \right)$, with unit vectors $\left( {\bm e}_r, {\bm e}_\theta, {\bm e}_{\phi \text{3D}} \right)$.
Both boundaries of the spherical shells are considered as electrically insulating, mechanically rigid, and we impose a fixed temperature contrast $\Delta T = T_i - T_o = T_\text{3D}(r_i) - T_\text{3D}(r_o)$ which drives convection.

In order to non-dimensionalise our variables, we use the shell thickness $d = s_o - s_i$ as the reference length-scale, the viscous diffusion time $d^2 / \nu$ as the reference time-scale, the temperature contrast between the boundaries $\Delta T$ as the reference for temperature, and $\sqrt{\rho \mu_0 \lambda \Omega}$ -- where $\rho$ and $\lambda$ are respectively the density and the magnetic diffusivity of the fluid and $\mu_0$ is the magnetic permeability of the vacuum -- as the reference for the magnetic field.
In such context, our system is controlled by four dimensionless parameters: the Ekman number, the Rayleigh number, the Prandtl number and the magnetic Prandtl number which are respectively defined by
\begin{align}
\label{eq:adim_par}
Ek = \dfrac{\nu}{\Omega d^2}\,, \; Ra = \dfrac{\alpha_T g_o \Delta T d^3}{\kappa \nu}\,, \; Pr = \dfrac{\nu}{\kappa}\,, \; Pm = \dfrac{\nu}{\lambda}\,,
\end{align}
where $\alpha_T$ is the thermal expansion coefficient, $g_o = g(r_o)$ is the gravity at the outer boundary, and $\kappa$ is the thermal diffusivity. 
Note that the magnetic Prandtl number can equally be thought of as the ratio between the magnetic diffusion time $\tau_\lambda$ and the viscous diffusion time $\tau_\nu$, {\it i.e.} $Pm = \tau_\lambda / \tau_\nu$.

We further assume that the dynamics is well described by the evolution of the axial vorticity averaged in the $z$ direction $\omega_z$, such that the dynamics is restricted to that in the equatorial plane of the spherical shell \citep[{\it e.g.},][]{maffei2017characterization}.
Thus, the horizontal components of the velocity field ${\bm u}_\perp$, perpendicular to the rotation axis, are assumed to be mostly invariant along the rotation axis, {\it i.e.} ${\bm u}_\perp \sim (u_s, u_\phi, 0)$, where $u_s$ and $u_\phi$ are respectively the radial and azimuthal velocities.
The axial velocity $u_z$ is considered as varying linearly with $z$ in the direction of the rotation axis, including mass conservation at the spherical outer boundary, and the Ekman pumping contribution ${\cal P}$ \citep{schaeffer2005quasigeostrophic,gastine2019pizza}, yields
\begin{align}
\label{eq:uz_linearity}
u_z(s,\phi,z) = z \left[ \beta u_s + \dfrac{Ek}{2} \mathcal{P}(Ek, {\bm u}_\perp, \omega_z) \right]\,,
\end{align}
where the Ekman pumping term ${\cal P}(Ek, {\bm u}_\perp, \omega_z)$ is deduced from Greenspan's formula \citep{greenspan1968theory} in a rigid sphere, {\it i.e.}
\begin{align}
\label{eq:Ek_pump}
\mathcal{P}(Ek, {\bm u}_\perp, \omega_z) = - \left(\dfrac{s_o}{Ek}\right)^{1/2} \dfrac{1}{h^{3/2}}\,\left[ \omega_z - \dfrac{\beta}{2} u_\phi + \beta \dfrac{\partial u_s}{\partial \phi} - \dfrac{5 s_o}{2h} u_s \right]\,,
\end{align}
with $\beta = \dfrac{1}{h}\dfrac{\mathrm{d} h}{\mathrm{d} s} = - \dfrac{s}{h^2}$, and $h \equiv \sqrt{s_o^2 - s^2}$, the half-height of a cylinder aligned with the rotation axis at a radius $s$.
Note that the singularity of $\beta$ at $s=s_o$ is not an issue since mechanical boundary conditions enforce ${\bm u} = {\bm 0}$ there.

In this framework, the continuity equation $\nabla \cdot {\bm u} = 0$ reads
\begin{align}
\label{eq:QG_approximation}
\dfrac{\partial (s u_s)}{\partial s} + \dfrac{\partial u_\phi}{\partial \phi} + \beta s u_s = 0\,,
\end{align}
from which it follows that there is a streamfunction $\psi$, which satisfies
\begin{align}
\label{eq:QG_streamfunct}
u_s = \dfrac{1}{s}\dfrac{\partial \psi}{\partial \phi}\,, \; \; u_\phi = \overline{u_\phi} - \dfrac{\partial \psi}{\partial s} - \beta \psi\,,
\end{align}
which accounts for the non-axisymmetric QG-velocity.
$\overline{u_\phi}$ is the remaining axisymmetric zonal flow component, with the overbar $\overline{x}$ denoting the azimuthal average of any quantity $x$, {\it i.e.}
\begin{align}
\label{eq:phi_average}
\overline{x} \equiv \dfrac{1}{2 \pi} \displaystyle\int_{0}^{2\pi}  x\, \mathrm{d}\phi\,.
\end{align}

The dynamics of the non-axisymmetric motions are then described by the time-evolution of the $z$-averaged axial vorticity $\omega_z \equiv \left< ({\bm \nabla} \times {\bm u}) \cdot {\bm e}_z \right>$, where the angular brackets $\langle x \rangle$ refer to the axial average of any quantity $x$, such that
\begin{align}
\label{eq:z_average}
\langle x \rangle \equiv \dfrac{1}{2h} \displaystyle\int_{-h}^{h}  x\, \mathrm{d}z\,.
\end{align}
The axial vorticity can be expressed in our framework as
\begin{align}
\label{eq:QG_vortz-psi}
\omega_z = \dfrac{1}{s}\dfrac{\partial (s \overline{u_\phi})}{\partial s} - \nabla^2 \psi - \dfrac{1}{s}\dfrac{\partial (\beta s \psi)}{\partial s}\,,
\end{align}
and its time evolution reads
\begin{align}
\label{eq:momentum_QG-hyb-mag}
\dfrac{\partial \omega_z}{\partial t} +{\bm \nabla}_\perp \cdot \left( {\bm u}_\perp \omega_z \right) - \dfrac{2}{Ek} \beta u_s = \nabla_\perp^2 \omega_z - \dfrac{Ra}{Pr} \left\langle \dfrac{1}{r_o} \dfrac{\partial \vartheta_\text{3D}}{\partial \phi_\text{3D}} \right\rangle \\
+ \dfrac{1}{Ek\,Pm} \left\langle {\bm \nabla} \times ( {\bf j} \times {\bf B}) \cdot {\bm e}_z \right\rangle \nonumber \\
+ \mathcal{P}(Ek, {\bm u}_\perp, \omega_z)\,, \nonumber
\end{align}
where the subscript $_\perp$ corresponds to the horizontal part of the operators and ${\bf j} \equiv {\bm \nabla} \times {\bf B}$.

Compared to the classical QG axial vorticity model, we have here followed the hybrid approach of \citet{guervilly2016subcritical} and \citet{barrois2022comparison} and used the full 3D temperature and magnetic fields.
The above equation (\ref{eq:momentum_QG-hyb-mag}) is thus coupled with the 3D temperature equation
\begin{align}
\label{eq:heat_3D}
\dfrac{\partial \vartheta_\text{3D}}{\partial t} + {\bm u}_\text{3D} \cdot {\bm \nabla} \vartheta_\text{3D} + u_r \dfrac{\mathrm{d} T_\text{3D}^\text{cond}}{\mathrm{d}r} =
\dfrac{1}{Pr} \nabla^2 \vartheta_\text{3D}\,,
\end{align}
and with the 3D magnetic induction equation
\begin{align}
\label{eq:induction_3D}
\dfrac{\partial {\bf B}}{\partial t} = {\bm \nabla} \times ({\bm u}_\text{3D} \times {\bf B})+ \dfrac{1}{Pm} \nabla^2 {\bf B}\,,
\end{align}
where ${\bm u}_\text{3D} = \left( u_r, u_\theta, u_{\phi \text{3D}} \right)$ is the 3D-velocity in spherical coordinates.

In the above equations $T_\text{3D}^\text{cond}$ is the conducting temperature profile, a solution of $\nabla^2 T_\text{3D}^\text{cond} = 0$.
For a fixed temperature contrast between $r_i$ and $r_o$ without internal heating this takes the form
\begin{align}
\label{eq:T_3D_cond_Dormy_3D}
T_\text{3D}^\text{cond}(r) = \dfrac{r_o r_i}{r} - r_i,
\; \dfrac{\mathrm{d} T_\text{3D}^\text{cond}}{\mathrm{d} r} = -\dfrac{r_i r_o}{r^2}\,.
\end{align}

We reconstruct the 3D-velocity field ${\bm u}_\text{3D}$ from the QG velocity field ${\bm u}_\perp$ using the conversion between cylindrical and spherical coordinate systems, where the cylindrical quantities are cast onto the 3D-grid using a bi-linear extrapolation \citep[see][Appendix~D for more details]{barrois2022comparison}, such that
\begin{equation}
\label{eq:vel_3D-interp}
\left\lbrace\begin{aligned}
u_r(r,\theta,\phi_\text{3D})& = \sin\theta\, u_s(s,\phi) + \cos\theta\, u_z(s,\phi,z)\,, \\
u_\theta(r,\theta,\phi_\text{3D}) &= \cos\theta\, u_s(s,\phi) - \sin\theta\, u_z(s,\phi,z)\,, \\
u_{\phi \text{3D}}(r,\theta,\phi_\text{3D})& = u_\phi(s,\phi) + {\cal T}_w(r,\theta)\,,
\end{aligned}\right.
\end{equation}
where $u_z$ is obtained from Eq.~(\ref{eq:uz_linearity}), and with an additional contribution to the axisymmetric azimuthal motions $\overline{u_{\phi \text{3D}}}$ of the thermal wind, ${\cal T}_w(r,\theta)$, which satisfies the relation
\begin{align}
\label{eq:thw-u_phi3D_formula}
\dfrac{\overline{\partial u_{\phi \text{3D}}}}{\partial z} = \dfrac{Ra\,Ek}{2 Pr} \dfrac{g(r)}{r} \dfrac{\partial \overline{\vartheta_\text{3D}}}{\partial \theta}\,,
\end{align}
and is integrated between the position $z$ and the height of the column above the equator $h$, {\it i.e.}
\begin{align}
\label{eq:thw-u_phi3D}
{\cal T}_w(r,\theta) = \dfrac{Ra Ek}{2 Pr} \int_{h}^{z} \dfrac{g(r)}{r}\dfrac{\partial \overline{T_\text{3D}}}{\partial \theta} \mathrm{d} z' \,,
\end{align}
where $g(r)=r/r_o$ is the dimensionless 3D gravity field.
Note that inside the tangent cylinder, apart from the thermal wind contribution, the 3D velocitiy components are set to zero and thus mainly temperature or magnetic diffusion occurs in that region.

Finally, the $z$-averaged axial vorticity equation (\ref{eq:momentum_QG-hyb-mag}) has to be supplemented by an equation to account for the axisymmetric motions, that is
\begin{align}
\label{eq:zonal_QG-mag}
\dfrac{\partial \overline{u_\phi}}{\partial t} + \displaystyle\overline{u_s \dfrac{\partial u_\phi}{\partial s}} + \displaystyle\overline{\dfrac{u_s u_\phi}{s}} = \nabla_\perp^2 \overline{u_\phi} - \dfrac{1}{s^2} \overline{u_\phi} \\
+ \dfrac{1}{Ek\,Pm} \left< \overline{({\bf j} \times {\bf B}) \cdot {\bm e}_\phi} \right> \nonumber \\
- \left(\dfrac{s_o}{Ek}\right)^{1/2} \dfrac{1}{h^{3/2}}\,\overline{u_\phi}\,, \nonumber
\end{align}
where the last term of the right-hand-side is the Ekman-pumping contribution to the axisymmetric motions.

\subsection{Computation of the QG-Lorentz Force}
\label{sec:QG-LF-comp}

As seen in equations (\ref{eq:momentum_QG-hyb-mag}) and (\ref{eq:zonal_QG-mag}), our method requires that we compute the following two quantities related to the 3D Lorentz force averaged over axial direction
\begin{align}
\label{eq:LF-omega_z}
{\bf F}_{{\cal L}\,,\,\omega_z} = \dfrac{1}{Ek\,Pm} \left< {\bm \nabla} \times ({\bf j} \times {\bf B}) \cdot {\bm e}_z \right>\,,
\end{align}
\begin{align}
\label{eq:LF-uphi_0}
{\bf F}_{{\cal L}\,,\,\overline{u_\phi}} = \dfrac{1}{Ek\,Pm} \left< \overline{({\bf j} \times {\bf B}) \cdot {\bm e}_\phi} \right>\,.
\end{align}

We can expand Eq.~(\ref{eq:LF-omega_z}) as
\begin{align}
\label{eq:LF-omega_z_curlz}
{\bf F}_{{\cal L}\,,\,\omega_z} = \dfrac{1}{Ek\,Pm} \left< \dfrac{1}{s} \left( \dfrac{\partial }{\partial s} \left[ s ({\bf j} \times {\bf B}) \cdot {\bm e}_\phi \right] - \dfrac{\partial }{\partial \phi} \left[ ({\bf j} \times {\bf B}) \cdot {\bm e}_s \right] \right) \right>\,,
\end{align}
and using the identity ${\bm e}_s = \sin \theta\, {\bm e}_r + \cos \theta\, {\bm e}_\theta$ this becomes
\begin{align}
\label{eq:LF-omega_z_3D}
{\bf F}_{{\cal L}\,,\,\omega_z} = \dfrac{1}{Ek\,Pm} \left< \dfrac{1}{s} \left( \dfrac{\partial }{\partial s} \left[ s ({\bf j} \times {\bf B}) \cdot {\bm e}_\phi \right] \right. \right. \\
\left. \left. - \dfrac{\partial }{\partial \phi} \left[ \sin \theta\, ({\bf j} \times {\bf B}) \cdot {\bm e}_r
+ \cos \theta\, ({\bf j} \times {\bf B}) \cdot {\bm e}_\theta \right] \right) \right>\,. \nonumber
\end{align}

Recalling that the $s$, $\phi$ and $z$ components are orthogonal, that the $z$-averaging operator is linear so $\left< u + v \right> = \left< u \right> + \left< v \right>$, and adopting the more compact notation ${\bf f} \cdot {\bm e}_x \equiv {\bf f}_x$ for any field ${\bf f}$ and coordinate $x$, the Lorentz force terms become
\begin{align}
\label{eq:LF-QG_3D}
{\bf F}_{{\cal L}} = \left\lbrace\begin{aligned}
&\dfrac{1}{Ek\,Pm} \left( \dfrac{1}{s} \left< \dfrac{\partial }{\partial s} \left[ s ({\bf j} \times {\bf B})_\phi \right] \right> \right. \\
&\hspace*{0.7cm} \left. - \dfrac{1}{s} \dfrac{\partial }{\partial \phi} \left[ \left< \sin \theta\, ({\bf j} \times {\bf B})_r + \cos \theta\, ({\bf j} \times {\bf B})_\theta \right> \right] \right) \\
&\dfrac{1}{Ek\,Pm} \left< \overline{({\bf j} \times {\bf B})_\phi} \right>
\end{aligned}\right.\,.
\end{align}

Regarding practical implementation, we find it useful to make use of the Leibniz's rule to switch the order of the $s$-derivative and $z$-integration steps in the first term of the curl in Eq.~(\ref{eq:LF-QG_3D}),
\begin{align}
\label{eq:LF-QG_3D-Leibniz}
{\bf F}_{{\cal L}\,,\,\omega_z} = \dfrac{1}{Ek\,Pm} \left( \dfrac{1}{s} \left[ \dfrac{\partial }{\partial s} \left[ s \left< ({\bf j} \times {\bf B})_\phi \right> \right] \right. \right. \\
\left. \left. + \beta s \left( \left< ({\bf j} \times {\bf B})_\phi \right> - \dfrac{1}{2} \left[ ({\bf j} \times {\bf B})_\phi \right](\pm h) \right) \right] \right. \nonumber \\
\left. - \dfrac{1}{s} \dfrac{\partial }{\partial \phi} \left[ \left< \sin \theta\, ({\bf j} \times {\bf B})_r + \cos \theta\, ({\bf j} \times {\bf B})_\theta \right> \right] \right)\,, \nonumber
\end{align}
where the surface term $({\bf j} \times {\bf B})_\phi(\pm h)$ cancels when ${\bf B}$ matches to a potential field at the outer boundary.
This avoids the need to compute the $s$-derivative on the 3D physical grid using a finite-difference scheme, before the $z$-averaging step.
Appendix~\ref{sec:LF-bench} presents validation and benchmark tests that have been carried out to verify our computations of these QG-Lorentz force terms.

\subsection{Numerics}
\label{sec:Numerics}

The calculations presented in this study have been carried out using an extension of the open-source pseudo-spectral spherical QG code {\tt pizza} \citep{gastine2019pizza,barrois2022comparison}, written in Fortran and freely available at \url{https://github.org/magic-sph/pizza/tree/hybrid_QG-3D} under the GNU GPL v3 license.
The 2D quantities are expanded in Fourier series up to the degree $N_m$ in the azimuthal direction and in Chebyshev polynomials up to degree $N_s$ in the radial direction.
The 3D fields are expanded in Spherical Harmonics up to the degree and order $\ell_\text{max}$ in the angular $(\theta,\phi_\text{3D})$ directions and in Chebyshev polynomials with $N_r$ collocation grid points in the radial direction.
The open-source \texttt{SHTns}\footnote{\url{https://bitbucket.org/nschaeff/shtns}} library is employed to handle the Spherical Harmonic Transforms \citep{schaeffer2013efficient}.
Parallelisation of the hybrid QG-3D code relies on the Message Passing Interface ({\tt MPI}) library.

The azimuthal (respectively radial) expansion involves adding zeros if $m_\text{3D} > m$ ($N_r > N_s$) and truncating fields at $m = m_\text{3D} = \ell$ ($N_s = N_r$) if $m_\text{3D} < m$ ($N_r < N_s$).
The same 3D and QG grids have been used when possible but because we explored cases with both $Pm < 1$ and $Pr <1$, on some occasions we allowed the grid size to vary between fields.
In these cases, nothing in particular has been done to bridge the two grids although we have used hyperdiffusion to mitigate this closure problem \citep{schaeffer2005thesis}.

\subsection{Hyperdiffusion}
\label{sec:Hdif}

We included in our implementation an option to use hyperdiffusion.
Following the formalism of \citet{nataf2015turbulence} and \citet{aubert2017spherical}, the diffusion operators entering Eqs.~(\ref{eq:momentum_QG-hyb-mag})-(\ref{eq:induction_3D})-(\ref{eq:zonal_QG-mag}) are multiplied by hyperdiffusivity functions that solely depend on the azimuthal wavenumber $m$ or on the spherical harmonic degree $\ell$, such that
\begin{align}
\label{eq:hdif-vel-2D}
f_{H, u}(m) = \left\lbrace\begin{aligned}
&1\, \; \hspace*{1cm} \text{for } m < m_H\,, \\
&q_{H, u}^{m-m_H}\, \; \text{for } m \geq m_H\,,
\end{aligned}\right.\,.
\end{align}
on the velocity field, and
\begin{align}
\label{eq:hdif-mag-JA}
f_{H, B}(\ell) = \left\lbrace\begin{aligned}
&1\, \; \hspace*{0.75cm} \text{for } \ell < \ell_H\,, \\
&q_{H, B}^{\ell-\ell_H}\, \; \text{for } \ell \geq \ell_H\,,
\end{aligned}\right.\,.
\end{align}
on the magnetic field, where $\ell_H$ and $m_H$ are the cut-off degrees and azimuthal orders below which the hyperdiffusion has no effect.
$q_{H, u}$ and $q_{H, B}$ are the strength of the hyperdiffusive effect on ${\bm u}_\perp$ or ${\bf B}$ respectively and have been varied in the range $1.01 \leq q_H \leq 1.08$ (we do not apply any hyperdiffusion to the temperature field).
The values for $m_H$ and $\ell_H$ for the runs using hyperdiffusion on either ${\bm u}_\perp$ or ${\bf B}$ are summarised in Table~\ref{tab:run_list}.

We have employed hyperdiffusion for two main reasons: ({\it i}) to mitigate closure issues when the 2D and 3D grids were different
; and ({\it ii}) to avoid the numerical problems in our most demanding runs -- {\it i.e.} with the highest $Rm$ -- arising because of the tangent cylinder discontinuity and the approximations involved in the interpolation schemes.
We have successfully removed the hyperdiffusion in several cases without observing any significant changes in the average properties.

\subsection{Diagnostics}
\label{sec:Posterior_diagnostic}

We now introduce the following notations for our various integral and average operators.
For any quantity $x$, the hat $\widehat{x}$ corresponds to its time average, that is
\begin{align}
\label{eq:time_average}
\widehat{x} \equiv \dfrac{1}{\tau}\displaystyle\int_{t_0}^{t_0+\tau} x\, \mathrm{d}t\,,
\end{align}
where $t_0$ is chosen such that any transient has been passed and with $\tau$ the averaging time interval long enough to reach a statistical equilibrium.
The brackets $\lbrace x \rbrace_\text{3D}$ corresponds to the full spherical shell average, the brackets $\lbrace x \rbrace_\text{S}$ to a spherical surface average and the brackets $\lbrace x \rbrace_\text{QG}$ to the average over the equatorial annulus, such that, respectively
\begin{align}
\label{eq:spatial_averages}
\lbrace x \rbrace_\text{3D} \equiv \dfrac{1}{{\cal V}_\text{3D}} \int_{\mathcal{V}_\text{3D}} x\, \mathrm{d} \mathcal{V}\,, \;
\lbrace x \rbrace_\text{S} \equiv \dfrac{1}{4\pi} \int_0^\pi \int_0^{2\pi} x\, \sin \theta\, \mathrm{d} \theta\, \mathrm{d} \phi_\text{3D}\,, \; \\
\lbrace x \rbrace_\text{QG} \equiv \dfrac{2}{{\cal V}_\text{QG}} \int_{\mathcal{V}_\text{QG}} x\, h(s) s\,\mathrm{d} s\,\mathrm{d} \phi\,, \nonumber
\end{align}
with $\mathcal{V}_\text{3D}$ the volume of the full spherical shell and $\mathcal{V}_\text{QG}$ the volume outside of the tangent cylinder.

The dimensionless kinetic energy per unit volume $E_\text{kin}$, is defined by 
\begin{align}
\label{eq:E_kin}
E_\text{kin} = \dfrac{1}{2} \lbrace {\bm u}_\perp^2 \rbrace_\text{QG}\,.
\end{align}
We similarly define the dimensionless magnetic energy per unit volume $E_\text{mag}$, as
\begin{align}
\label{eq:E_mag}
E_\text{mag} = \dfrac{1}{2} \lbrace {\bf B}^2 \rbrace_\text{3D}\,.
\end{align}
The magnetic to kinetic energy ratio is then
\begin{align}
\label{eq:EkinEmag_ratio}
{\cal M} = \dfrac{\widehat{E}_\text{mag}}{\widehat{E}_\text{kin}}\,.
\end{align}

From these expressions, we define a diagnostic for the fluid velocity which characterises the average flow speed, based on the root-mean-square (r.m.s.) of the velocity, and which is denoted by the Reynolds number
\begin{align}
\label{eq:Reynolds}
Re = \displaystyle\widehat{\sqrt{2 E_\text{kin}}}\,.
\end{align}
Then the magnetic Reynolds number is simply $Rm = Pm\,Re\,$ and we can finally define the Elsasser number, a non-dimensional measure of the magnetic field strength, which reads
\begin{align}
\label{eq:Elsasser}
\Lambda = \displaystyle\dfrac{Ek}{Pm}\,Rm^2\,{\cal M} = 2 Ek\,Pm\,\widehat{E}_\text{mag}\,.
\end{align}

\section{Results}
\label{sec:Results}

We present here results of experiments conducted at control parameters of $Ek=3 \times 10^{-5}$, $Pr=10^{-1}$, varying $Pm$, and focusing on a regime well above the onset of convection, {\it i.e.} $Ra \geqslant 5 \, Ra_c$.
Time-integrating the nonlinear equations  in absence of a magnetic field \citep{barrois2022comparison}, we estimated the critical Rayleigh number for this configuration to be $Ra_c = 1.03 \times 10^6$ (which is thus the thermal convection critical value).
In total 33 numerical simulations have been carried out, a list of their key diagnostics is given in Table~\ref{tab:run_list} in Appendix~\ref{sec:Append-A-Results}.
Our first successful dynamo case -- at parameters $Ek = 3 \times 10^{-5}$, $Pr = 0.1$, $Pm = 0.9$, $Ra = 1.66 \times 10^{7} \sim 16\,Ra_c$ -- was started from a motionless fluid with a strong axial dipole with $\Lambda \sim {\cal O}(1)$ and a random perturbation in temperature.
Subsequent experiments were initialized starting from a previously converged simulation.
Attempts to restart the configurations with the largest Elsasser numbers from a strong axial dipole state were found to again yield the same final weak field multipolar solution.

\subsection{Dynamo regime diagrams}
\label{sec:res-pizza-dyn-diag}

\begin{figure*}
\centering{
	\includegraphics[width=0.98\linewidth]{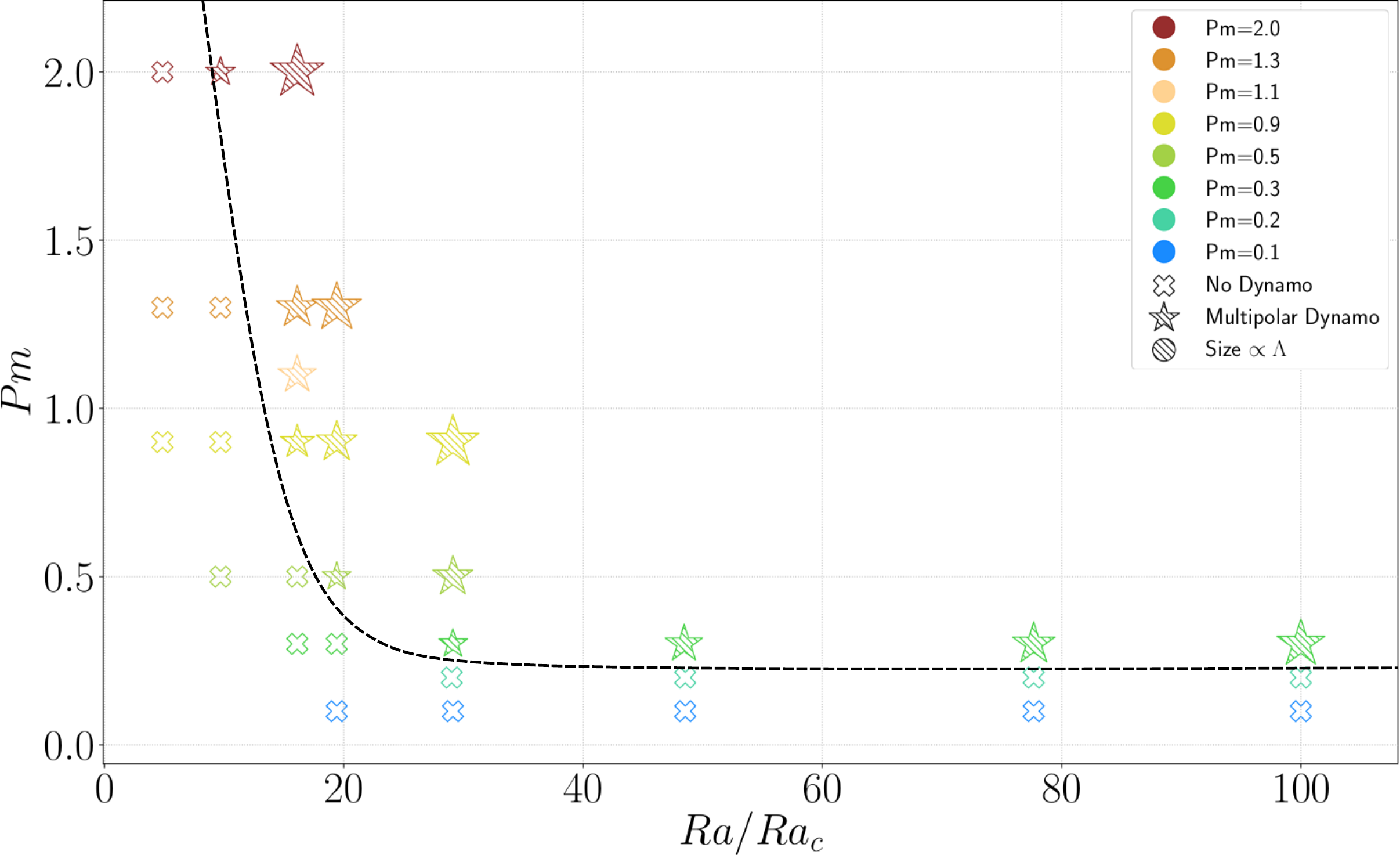}}
	\caption{
	Magnetic Prandtl number, $Pm$, as a function of the supercriticality, $Ra / Ra_c$.
	Dynamo regime diagram computed for a series of cases at $Ek = 3 \times 10^{-5}$ and $Pr=0.1$.
	Experiments that failed to produce a self-sustain dynamo are marked with a cross, those with a self-sustained multipolar dynamo are marked with a star and their symbol size is proportional to the Elsasser number, $\Lambda$.
    The different colors correspond to different $Pm$.
    The dashed line marks the tentative limit between failed and growing dynamos.
	}
	\label{fig:Summary-Laws_Pm-vs-Ra|Rac}
\end{figure*}

A dynamo regime diagram as a function of $Pm$ and $Ra/Ra_c$ for all our runs conducted at $Ek=3 \times 10 ^{-5}$, $Pr=0.1$ is presented in Figure~\ref{fig:Summary-Laws_Pm-vs-Ra|Rac}.
The crosses correspond to simulations which failed to produce a self-sustained dynamo while the stars represent the growing dynamos.
Similar to the case for 3D numerical dynamos we find that passing the onset for the dynamo action requires increasingly large $Pm$ on decreasing supercriticality.
The shape of the dynamo threshold found in this diagram is qualitatively similar to that found by \citep[][see their~Fig.1]{christensen2006scaling} despite the fact that they considered $Pr=1$ while we consider $Pr=0.1$.
An important difference to note though is that we only find multipolar dynamos with the hybrid QG-3D formalism.

All the dynamos we have found so far have a low magnetic to kinetic energy ratio ${\cal M} < 1$ and most of them have ${\cal M} < 10^{-1}$ (with the exception of some of the points, {\it e.g.} at $Pm=2.0$, $Ra/Ra_c \sim 16.1$ that reached a moderate ${\cal M}=0.15$ despite having the highest $\Lambda = 8.02$).
They therefore fall into the weak-field dynamo regime, characterised by ${\cal M} \ll 1$ \citep[{\it e.g.},][]{schaeffer2017turbulent,aubert2017spherical,schwaiger2019force}.

\begin{figure*}
\includegraphics[width=0.98\linewidth]{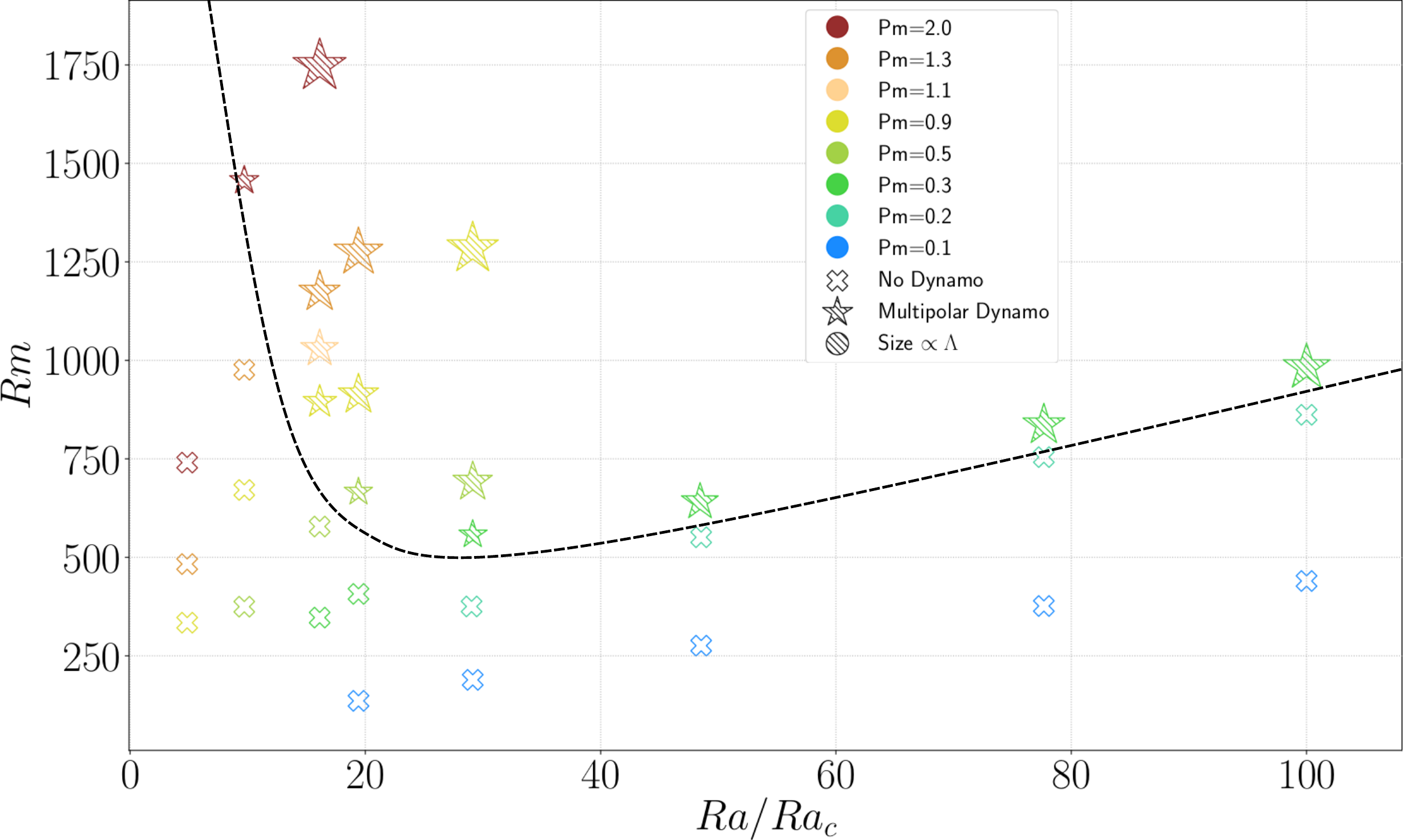}
	\caption{
	Magnetic Reynolds $Rm$ as a function of the supercriticality $Ra/Ra_c$.
	The symbols and the line have the same signification as in Fig.~\ref{fig:Summary-Laws_Pm-vs-Ra|Rac}.
	}
	\label{fig:Summary-Laws_Rm-Elsasser}
\end{figure*}

Plotting the magnetic Reynolds number $Rm$ against the supercriticality $Ra/Ra_c$ in Figure~\ref{fig:Summary-Laws_Rm-Elsasser}, we observe that the minimum $Rm$ required to obtain a self sustained dynamo in this setup is at least $500$, in agreement with previous QG magnetic studies \citep[{\it e.g.},][]{guervilly2010thesis}.
This value is about one order of magnitude higher than that found for 3D simulations, with for example, a critical value of $Rm_c \sim 50$ reported by \citet{petitdemange2018systematic}.
We found that $\Lambda > 0.05$ for all the cases when a self-sustained dynamo was identified -- with the highest value $\Lambda \sim 8$ reached for the case at $Ek = 3 \times 10^{-5}$, $Pr = 0.1$, $Pm = 2.0$, $Ra = 1.66 \times 10^{7} \sim 16.1\,Ra_c$ -- and we generally observe an increase of $\Lambda$ with increasing $Rm$.

\subsection{An example weak field dynamo}
\label{sec:res-Ek3e-5_Pr-1}

To illustrate the typical features of our data set, we look at an example dynamo with parameters $Ek = 3 \times 10^{-5}$, $Pr = 0.1$, $Pm = 0.9$, $Ra = 1.66 \times 10^{7} \sim 16\,Ra_c$ which was computed with a resolution of $(N_s , N_m)/(N_r, \ell_\text{max}) = (385, 768)/(149, 148)$ for approximately $\sim 5 \tau_\nu = 4.5 \tau_\lambda$.
This case is rather close to the onset of the dynamo action and decreasing $Pm$ by a factor $2$ or decreasing $Ra$ by a factor $1.5$ was sufficient to lose dynamo action (see Fig.~\ref{fig:Summary-Laws_Pm-vs-Ra|Rac}).
The average magnetic Reynolds number of this simulation is $Rm \simeq 900$ and ${\cal M} = 0.04$.

\begin{figure*}
\centering{
	\includegraphics[width=0.46\linewidth]{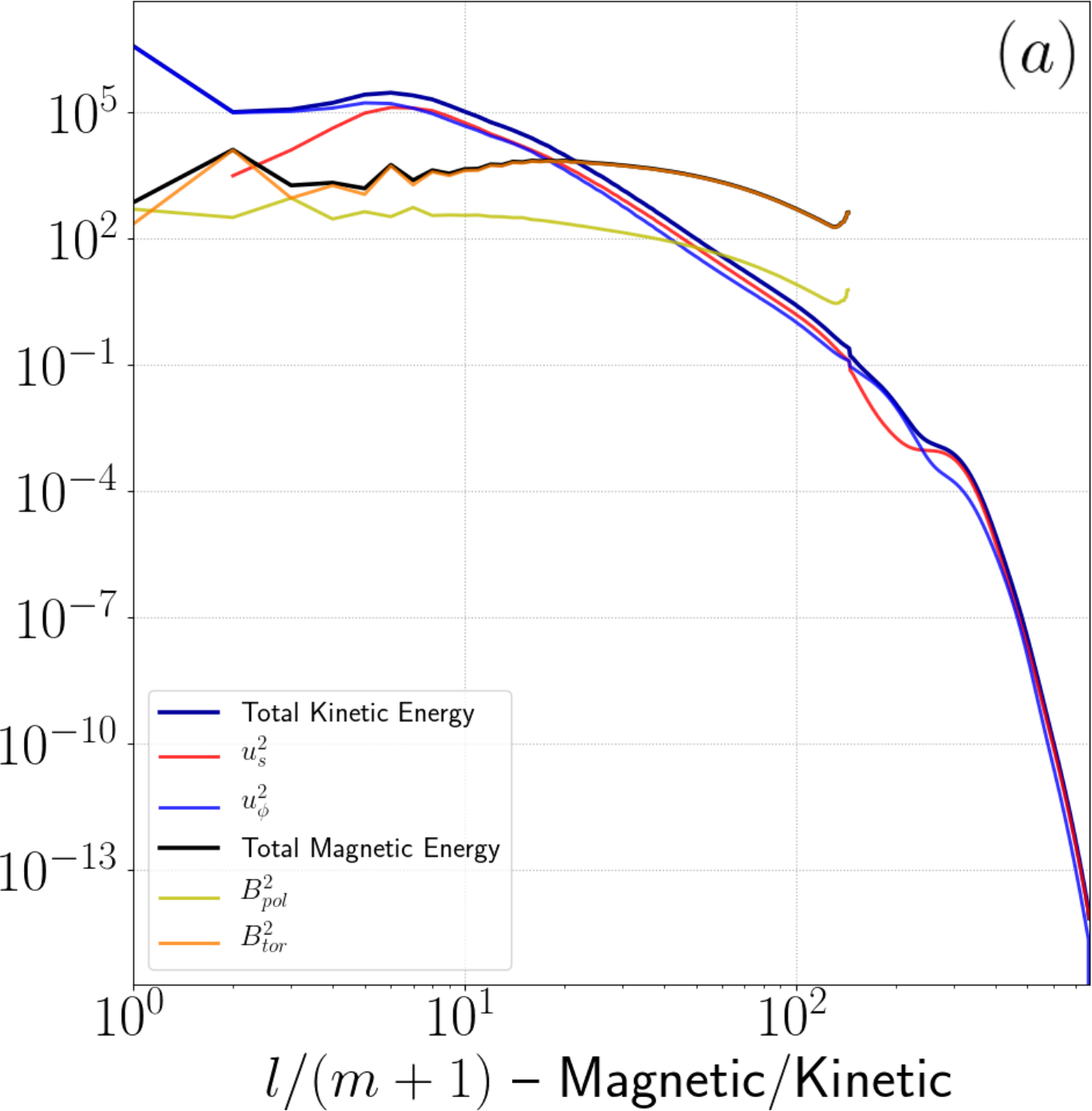}
	\includegraphics[width=0.53\linewidth]{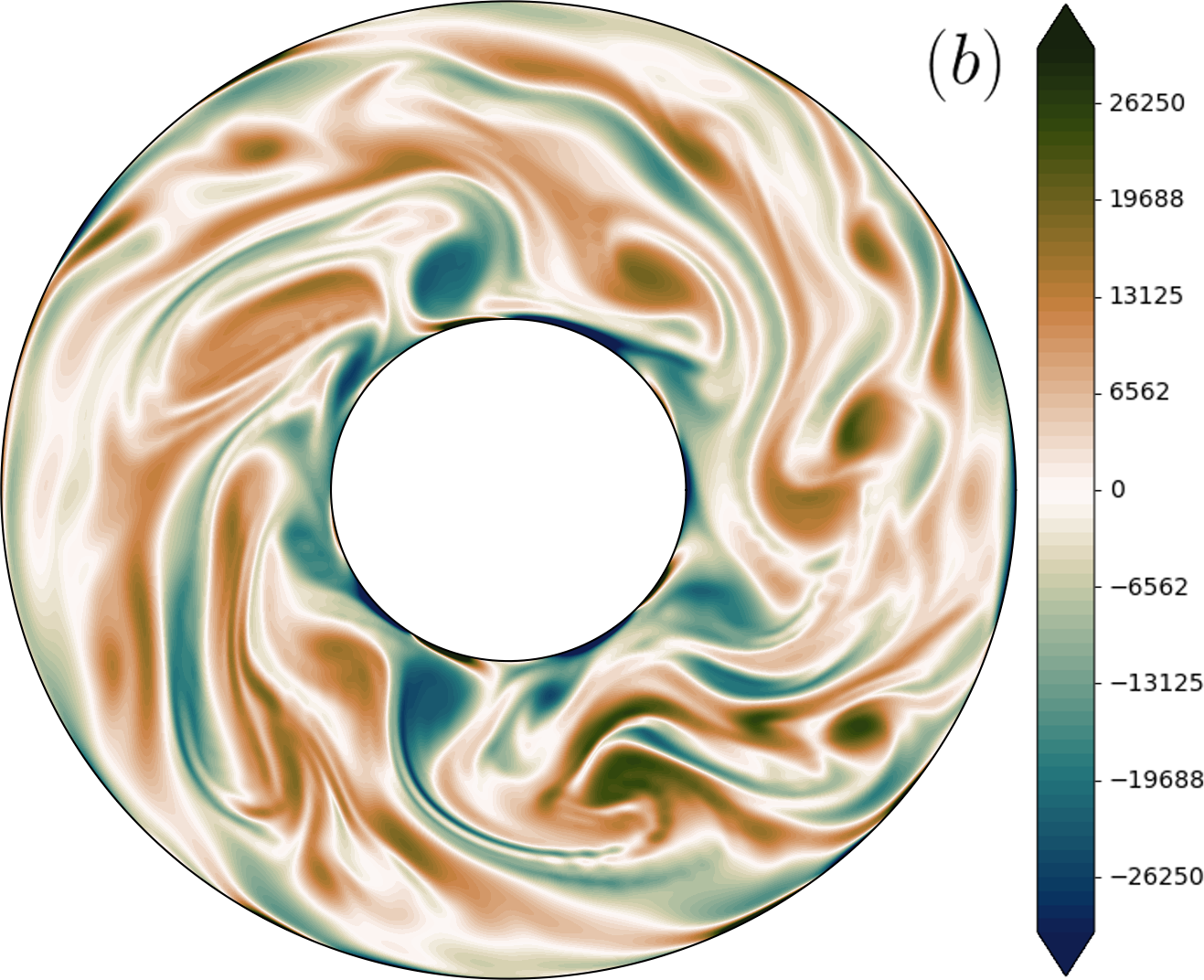}}
\centering{
	\includegraphics[width=0.19\linewidth]{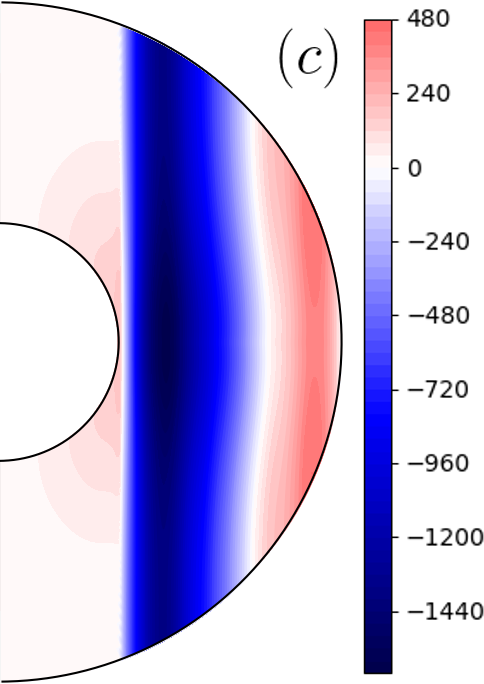}
	\includegraphics[width=0.59\linewidth]{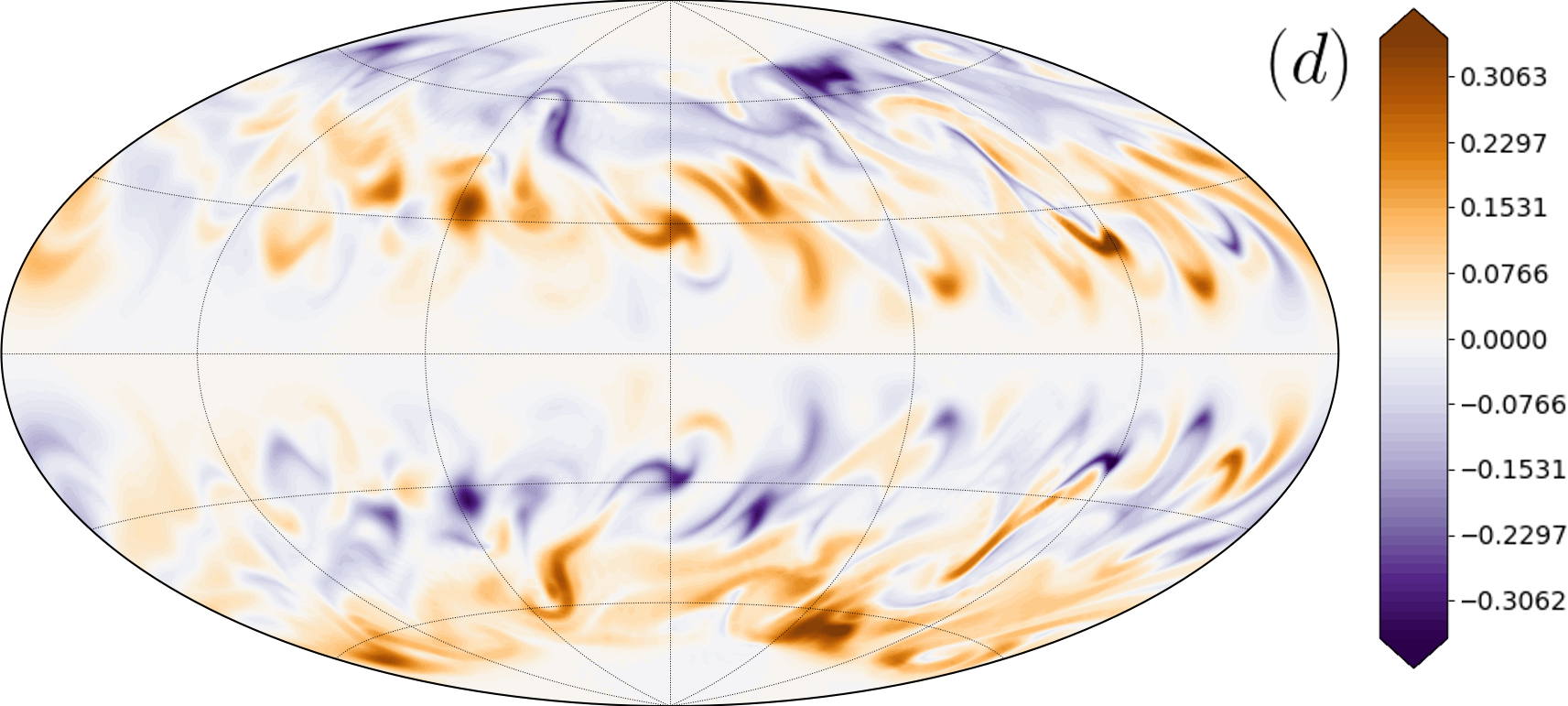}
	\includegraphics[width=0.19\linewidth]{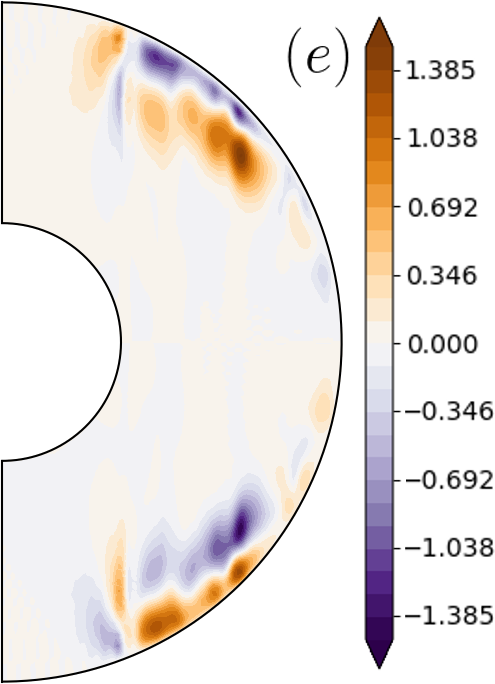}}
	\caption{
	(a) Time-averaged magnetic and kinetic energy spectra.
    (b) Snapshot of the $z$-averaged vorticity $\omega_z$.
    (c) Meridional section of the $\phi$-averaged azimuthal 3D velocity $\overline{u_{\phi \text{3D}}}$.
    (d) Snapshot of the radial magnetic field at the outer boundary $B_r(r_o)$.
    (e) Meridional section of the $\phi$-averaged azimuthal magnetic field $\overline{B_\phi}$.
    For a dynamo with control parameters $Ek = 3 \times 10^{-5}$, $Pr = 0.1$, $Pm = 0.9$, $Ra = 1.66 \times 10^{7} \sim 16\,Ra_c$.
	}
	\label{fig:Pm0o9_Ra15_Ekin-Emag}
\end{figure*}

Figure~\ref{fig:Pm0o9_Ra15_Ekin-Emag} displays the time-averaged magnetic and kinetic energy spectra (a), a snapshot of the $z$-averaged axial vorticity $\omega_z$ (b), a meridional section of the longitudinally-averaged azimuthal 3D velocity $\overline{u_{\phi \text{3D}}}$ (c), a snapshot of the radial magnetic field at the outer boundary of the dynamo region $B_r(r_o)$ (d), and a meridional section of the $\phi$-averaged azimuthal magnetic field $\overline{B_\phi}$ (e).

The power spectra shown in Fig.~\ref{fig:Pm0o9_Ra15_Ekin-Emag}~(a) confirms that the dynamo is multipolar (the magnetic field is dominated by the degree $\ell=2$) and that the kinetic energy dominates over the magnetic energy at the largest length-scales by about $2$ orders of magnitude (both for $\ell$ and $m+1 < 20$, where $m$ is here the QG azimuthal wavenumber).
Additionally, we can observe that the velocity field is dominated by the azimuthal velocity at large length scales ($m<8$) and that the magnetic field is strongly dominated by its toroidal part, this latter component being greater than the poloidal field by one order of magnitude at almost all degrees.
This prevalence of the kinetic energy over the magnetic energy and of the toroidal component over the poloidal component of the magnetic field (dominated by the degree $\ell = 2$) is typical of all our dynamos.

We can see in Fig.~\ref{fig:Pm0o9_Ra15_Ekin-Emag}~(b-c) that the vorticity field (b) and the azimuthal velocity fields (c) are similar to what can be observed for a non-magnetic simulation \citep[see, {\it e.g.} Fig.~3-4 in][]{barrois2022comparison} with azimuthally elongated structures on the scale of the container in the case of the vorticity and with a strong zonal flow in the case of $u_{\phi \text{3D}}$.
This suggests that the magnetic field does not strongly influence the velocity field, consistent with the low value of ${\cal M}$ for this case.

Turning to the magnetic field components, we can see in Fig.~\ref{fig:Pm0o9_Ra15_Ekin-Emag}~(d-e) that $B_r(r=r_o)$ (d) and $\overline{B_\phi}$ (e) have fairly low amplitude, and are mostly non-dipolar despite the clear equatorial anti-symmetry.
$B_r$ is dominated by small length-scales and the location of the strongest azimuthal magnetic field suggests that dynamo action mostly occurs at mid-latitudes and in the outer third of the shell.
One could suspect that Ekman pumping is an important contributor to dynamo action, since at $Ek = 3 \times 10^{-5}$ it is expected to have some impact close to the outer boundary.
However, although Ekman pumping has been shown to contribute to dynamo action close to the onset of dynamo \citep{busse1975model} it has been found in similar, but mechanically-forced, QG-models that removing the Ekman pumping flow does not significantly modify the dynamo onset \citep{schaeffer2006quasi}.
We investigated the role of Ekman pumping in this dynamo by removing the Ekman pumping contribution to ${\bm u}_\text{3D}$ -- second part of Eq.~(\ref{eq:uz_linearity}) -- used in the magnetic induction equation, and did not observe any major modifications in the resulting fields.
We also conducted a test removing the thermal wind contribution to $\overline{u_{\phi \text{3D}}}$ -- see Eq.~(\ref{eq:thw-u_phi3D_formula}) -- and again did not observe any major change in the dynamo action, in agreement with \citet{guervilly2010thesis} who also found that the thermal wind does not seem to have a strong impact on the dynamo onset.
Our results are therefore consistent with QG flows influenced by a $\beta$-effect in spherical geometry -- the first term of Eq.~(\ref{eq:uz_linearity}) -- being sufficient to sustain a dynamo, in agreement with the earlier findings of \citet{schaeffer2006quasi} and \citet{guervilly2010thesis}.

The magnetic field structure reported in Fig.~\ref{fig:Pm0o9_Ra15_Ekin-Emag} is typical of our results.
All our simulations display similar equatorially-antisymmetric, predominantly degree $2$, magnetic fields strongest at mid-to-high latitudes in the outer part of the shell.
Note that the symmetry (or more correctly the anti-symmetry) of our magnetic field remains the same as that of our initial field, as there is, by construction, no ingredient to break the symmetry in our QG flows.

\subsection{Dynamo mechanism}
\label{sec:res-dyn-meca}

\begin{figure*}
\centering{
	\includegraphics[width=0.32\linewidth]{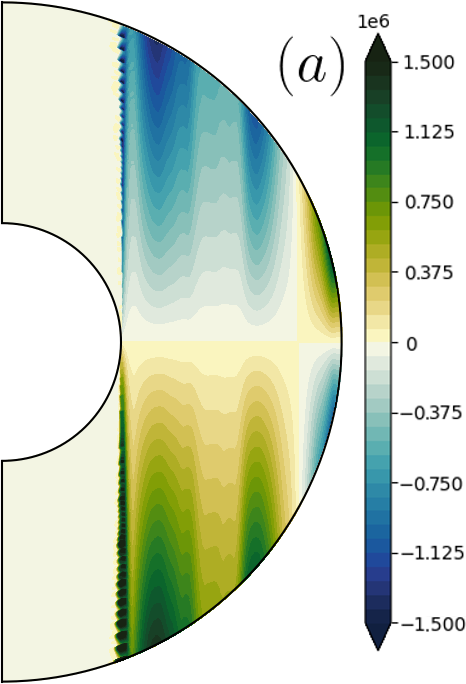}
	\includegraphics[width=0.32\linewidth]{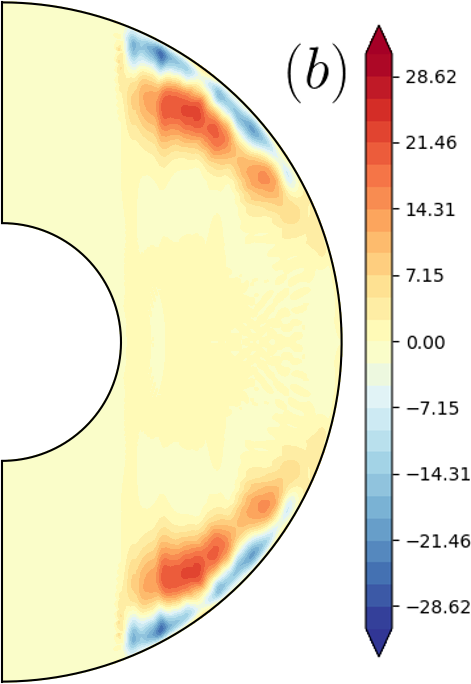}
	\includegraphics[width=0.32\linewidth]{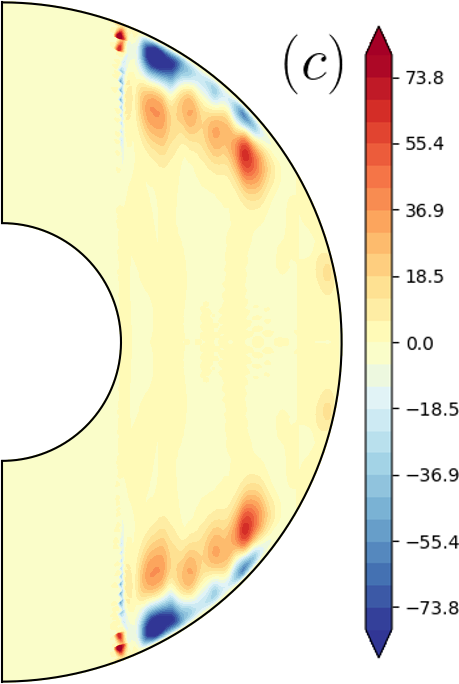}}
	\caption{
	(a) Meridional section of the longitudinally-averaged 3D-helicity $\overline{{\bm u}_\text{3D} \cdot \nabla \times {\bm u}_\text{3D}}$.
    (b) Azimuthal component of the mean electromotive force $\overline{({\bm u}_\text{3D}' \times {\bf B}') \cdot {\bm e}_\phi}$.
    (c) Estimated mean $\alpha$-effect $- \overline{\dfrac{d}{3\,\lbrace {\bm u}_\text{3D}' \rbrace_\text{S}} B_\phi \cdot ({\bm u}_\text{3D}' \cdot \nabla \times {\bm u}_\text{3D}')}$.
    For the same case as in Fig.~\ref{fig:Pm0o9_Ra15_Ekin-Emag}.
	}
	\label{fig:Pm0o9_Ra15_Hel+EMF+Alpha}
\end{figure*}

To illustrate the main mechanism underlying our dynamos, Figure~\ref{fig:Pm0o9_Ra15_Hel+EMF+Alpha} shows, from left to right respectively, meridional sections of the $\phi$-averaged 3D-helicity ${\cal H} \equiv \overline{{\bm u}_\text{3D} \cdot (\nabla \times {\bm u}_\text{3D})}$, the azimuthal component of the mean electromotive force $\overline{({\bm u}_\text{3D}' \times {\bf B}') \cdot {\bm e}_\phi}$ (hereafter EMF) and an estimate of the (rescaled) mean $\alpha$-effect $- \overline{\dfrac{\tau_c}{3} B_\phi \cdot ({\bm u}_\text{3D}' \cdot \nabla \times {\bm u}_\text{3D}')}$ for the same dynamo presented in the previous section -- where the prime $x'$ denotes fluctuations about the azimuthal average of any quantity $x$, {\it i.e.} $x' \equiv x - \overline{x}$ and where $\tau_c$ is a typical convective turnover time \citep{brandenburg2005astrophysical} which can be approximated by $\tau_c \sim d/\lbrace {\bm u}_\text{3D}' \rbrace_\text{S}$ following \citet{brown2010persistent} and \citet{gastine2012dipolar}.
We find that the mean helicity, often thought to be a key ingredient in the magnetic induction {\it via} the so-called $\alpha$-effect \citep[{\it e.g.},][]{moffatt1978magnetic,jones2008course}, changes sign between the North and South hemispheres and is stronger at mid latitudes and towards the outer boundary, where concentrations of $\overline{B_\phi}$ and strong zonal wind $\overline{u_{\phi \text{3D}}}$ have already been observed in the Fig.~\ref{fig:Pm0o9_Ra15_Ekin-Emag}.
Most of the 3D helicity is contained in its $z$-component ($u_z \cdot (\nabla \times {\bm u}_\text{3D}) \cdot {\bm e}_z$) with only weak contributions from the $s$- and $\phi$-components (not shown).
There is clearly a strong correlation between the mean EMF and the estimated $\alpha$-effect despite the discrepancy in amplitudes (Fig.~\ref{fig:Pm0o9_Ra15_Hel+EMF+Alpha}~b-c) with both quantities located where the maximum helicity and toroidal field are found, indicating that 
({\it i}) our model produces an $\alpha$-effect sufficient to support the dynamo action, 
({\it ii}) kinetic helicity provides a reasonable mean-field approximation of the actual EMF.

\begin{figure*}
\centering{
	\includegraphics[width=0.98\linewidth]{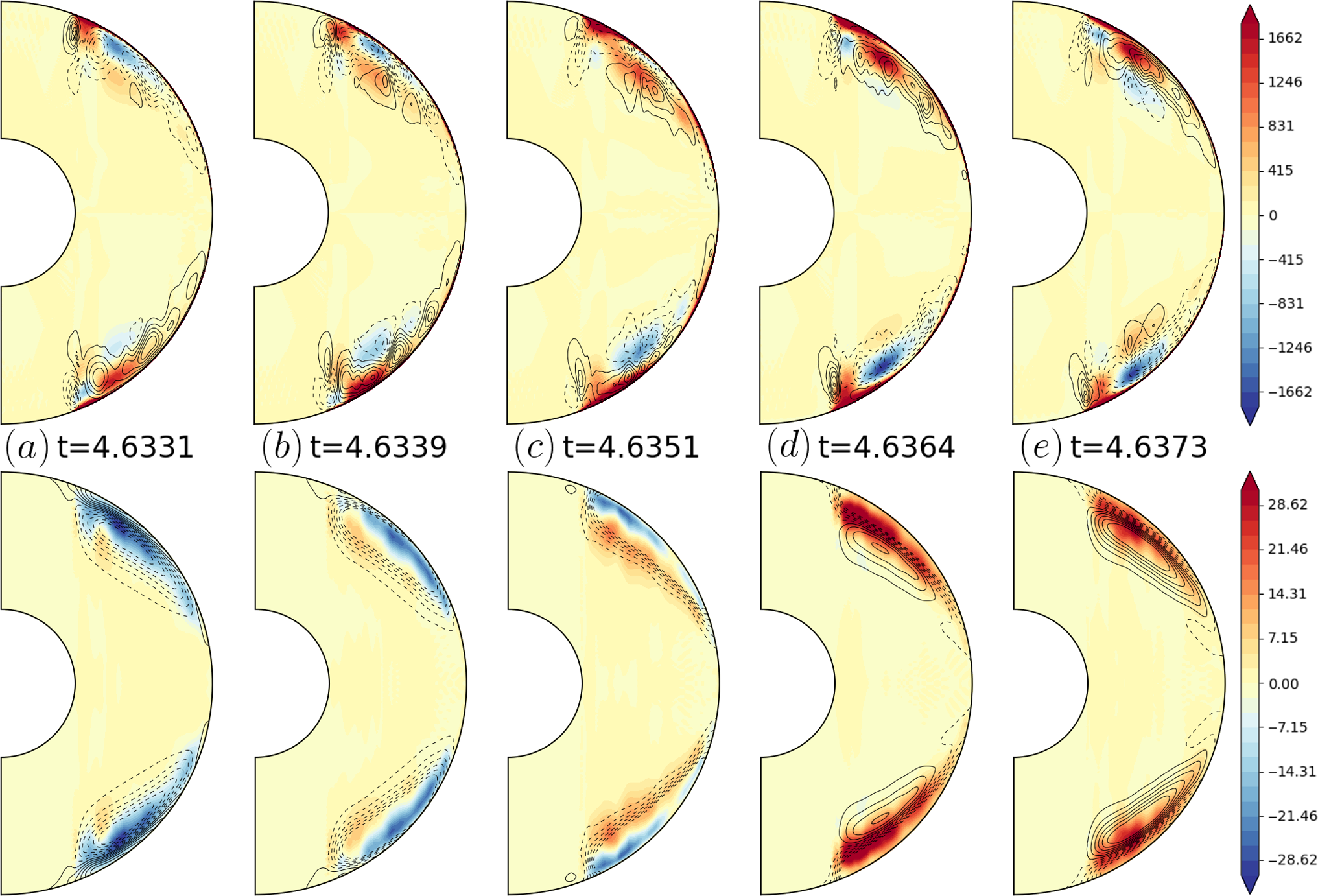}}
	\caption{
	Temporal evolution of $\phi$-averaged snapshots of the $\Omega$-effect $s\overline{\bf B}\cdot {\bm \nabla} \left(\dfrac{\overline{u_{\phi \text{3D}}}}{s}\right)$ with superimposed toroidal field lines (top row) and of the azimuthal component of the mean EMF $\overline{({\bm u}_\text{3D}' \times {\bf B}') \cdot {\bm e}_\phi}$ with poloidal field lines superimposed (bottom row) during half a cycle of a dynamo wave (from a to e).
    Note that the time difference between two snapshots is not constant and that time is expressed in units of the viscous diffusion time.
	}
	\label{fig:Pm0o9_Ra15_Om+TorLines_EMF+PolLines}
\end{figure*}

Further insight into the dynamo mechanism at work in this dynamo is provided by Fig.~\ref{fig:Pm0o9_Ra15_Om+TorLines_EMF+PolLines} which displays a sequence of longitudinally-averaged snapshots of the $\Omega$-effect $s\overline{\bf B}\cdot {\bm \nabla} \left(\dfrac{\overline{u_{\phi \text{3D}}}}{s}\right)$ \citep[{\it e.g.},][]{roberts2013genesis} with superimposed toroidal field lines (top row) and of the azimuthal component of the mean EMF $\overline{({\bm u}_\text{3D}' \times {\bf B}') \cdot {\bm e}_\phi}$ with superimposed poloidal field lines (bottom row).
A strong correlation is found between both the $\Omega$-effect and the toroidal field lines and between the EMF and the poloidal field lines, and both effects are concentrated in the same region localised in the upper part of the shell.
The $\Omega$-effect is stronger than the $\alpha$-effect by $1$-to-$2$ orders of magnitude and reaches its maxima in the region near to the outer boundary where both strong zonal flow and helicity are found.
This is not unexpected as strong zonal winds sustained by Reynolds stresses -- {\it i.e.} the correlations between the azimuthal motions $u_{\phi \text{3D}}'$ and the cylindrically radial velocity $u_s'$ -- are expected to produce an important $\Omega$-effect leading to multipolar dynamos dominated by the toroidal magnetic field.
This has been found in a number of previous studies: for classical Boussinesq models \citep{sheyko2016magnetic}, models with stress-free boundary conditions \citep{grote2000hemispherical,goudard2008relations,schrinner2012dipole}, models driven by strongly heterogeneous boundary heat flux \citep{dietrich2013hemispherical} and anelastic dynamo models \citep{gastine2012dipolar}.
The main mechanism of our dynamos can thus be understood as powered by a strong $\Omega$-effect -- generated by the strong shear of of the zonal flow -- with the toroidal field being converted into a poloidal field {\it via} an $\alpha$-effect \citep{parker1955hydromagnetic}.
Such multipolar dynamos are often classified as $\alpha \Omega$ or $\alpha^2 \Omega$ type, following the mean field nomenclature of {\it e.g.} \citet{steenbeck1966berechnung,steenbeck1969dynamo} or \citet{malkus1975macrodynamics}.

\subsection{Dynamo waves}
\label{sec:res-parker-waves}

Figure~\ref{fig:Pm0o9_Ra15_Om+TorLines_EMF+PolLines} also shows that within a small fraction of a viscous diffusion time the toroidal and poloidal field lines have been mostly pushed towards the poles and replaced by field lines of the reversed polarity while deeper in the bulk the underlying $\Omega$-effect and the EMF have changed sign.
This process is reminiscent of dynamo waves that have been found in a 3D dynamo models with a range of different geometries, boundary conditions and driving mechanisms \citep[see, {\it e.g.}][]{goudard2008relations,schrinner2011oscillatory,simitev2012far,sheyko2016magnetic} as well as in mechanically-forced QG dynamos \citep{schaeffer2006quasi}.

\begin{figure*}
\centering{
	\includegraphics[width=0.92\linewidth]{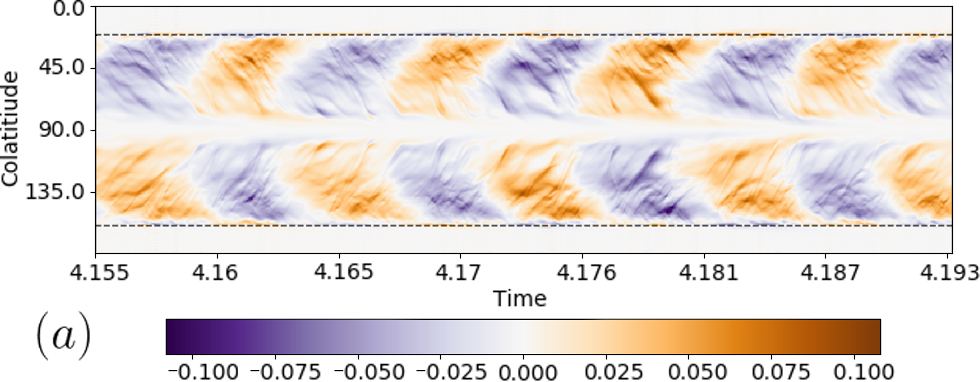}}
\centering{
	\includegraphics[width=0.92\linewidth]{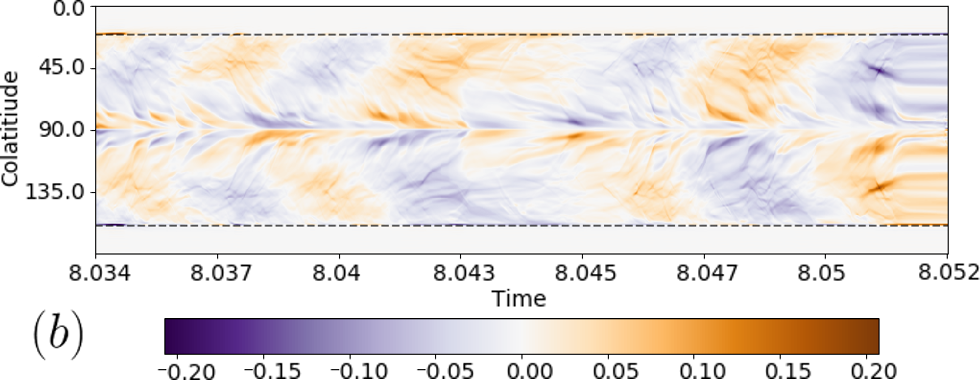}}
	\caption{
	Axisymmetric azimuthal magnetic field $\overline{B_\phi} (\theta)$ at $r \sim 0.92\,r_o$ as a function of time -- expressed in viscous units $\tau_\nu$ -- (so called butterfly diagram) for 
    (a) dynamo with control parameters $Ek = 3 \times 10^{-5}$, $Pr = 0.1$, $Pm = 0.9$, $Ra = 1.66 \times 10^{7}$ and,
    (b) a second case, more strongly forced, with control parameters $Ek = 3 \times 10^{-5}$, $Pr = 0.1$, $Pm = 0.9$, $Ra = 3.0 \times 10^7$.
    In both panels, dashed lines mark the location of the tangent cylinder.
	}
	\label{fig:Buterfly_Pm0o9_Ra15-30_BphiCMB}
\end{figure*}

In Figure~\ref{fig:Buterfly_Pm0o9_Ra15-30_BphiCMB} we show the temporal evolution of $\overline{B_\phi} (\theta)$ just beneath the outer boundary at $r \sim 0.92\,r_o$ -- a so-called “butterfly-diagram” -- for a case at $Ek = 3 \times 10^{-5}$, $Pr = 0.1$, $Pm = 0.9$, $Ra = 1.66 \times 10^{7} \sim 16\,Ra_c$ with a $Rm$ of $909$ (a) -- same dynamo as presented in \S\ref{sec:res-Ek3e-5_Pr-1} -- along with a more strongly driven case at $Ek = 3 \times 10^{-5}$, $Pr = 0.1$, $Pm = 0.9$, $Ra = 3.0 \times 10^7 \sim 30\,Ra_c$ with a higher $Rm$ of $1284$ (b).
In both cases, we observe that the flux patches of the toroidal magnetic field appear symmetrically in both hemispheres at low-to-mid latitudes and move towards the poles until they reach the tangent cylinder which inhibits further motions.
This is similar to what has been observed by {\it e.g.} \citet{schrinner2011oscillatory,gastine2012dipolar} and \citet{dietrich2013hemispherical}.
We find that the cycle starts again after the flux patches have travelled all the way to the tangent cylinder and we can also see that some of the toroidal field travels equatorwards and quickly vanishes.
The direction of migration (poleward or equatorward) of dynamo waves is controlled by gradients in the zonal flow \citep{yoshimura1975solar}.
Waves are expected to travel polewards when the zonal flow increases with cylindrical radius \citep[{\it e.g.},][]{sheyko2016magnetic}, consistent with the zonal flow patterns in our dynamos (see Fig.~\ref{fig:Pm0o9_Ra15_Ekin-Emag}~c).
On increasing $Rm$ between cases (a) and (b) the cycles become less coherent and less periodic with an oscillation period of $1$ cycle per $4.5 \times 10^{-3} \tau_\lambda$ and $1$ cycle per $2.7 \times 10^{-3} \tau_\lambda$ respectively.
This is consistent with the expectation that the reversal frequency should increase with increasing forcing or increasing field strength; we have increased $Ra/Ra_c$ leading to an increase in both $Rm$ and the magnetic field strength -- between cases (a) and (b).

In order to better characterise these oscillations, we test the approximate dispersion relation for (Parker) dynamo waves derived in \citet{schrinner2011oscillatory}, which reads
\begin{align}
\label{eq:Parker_wave_number}
\omega \sim \left( \dfrac{{\cal H}_\text{Int}'}{2\,r_o} \dfrac{Re_z}{Re_c} \right)^{1/2}\,,
\end{align}
with $Re_z = \sqrt{\lbrace \overline{u_{\phi \text{3D}}}^2 \rbrace_\text{3D}}$, $Re_c = \sqrt{\lbrace ({\bm u}_\text{3D}')^2 \rbrace_\text{3D}}$ and ${\cal H}_\text{Int}' = -\dfrac{1}{3} \lbrace |{\bm u}_\text{3D}' \cdot \nabla \times {\bm u}_\text{3D}'| \rbrace_\text{3D}$.

\begin{figure*}
\centering{
	\includegraphics[width=0.98\linewidth]{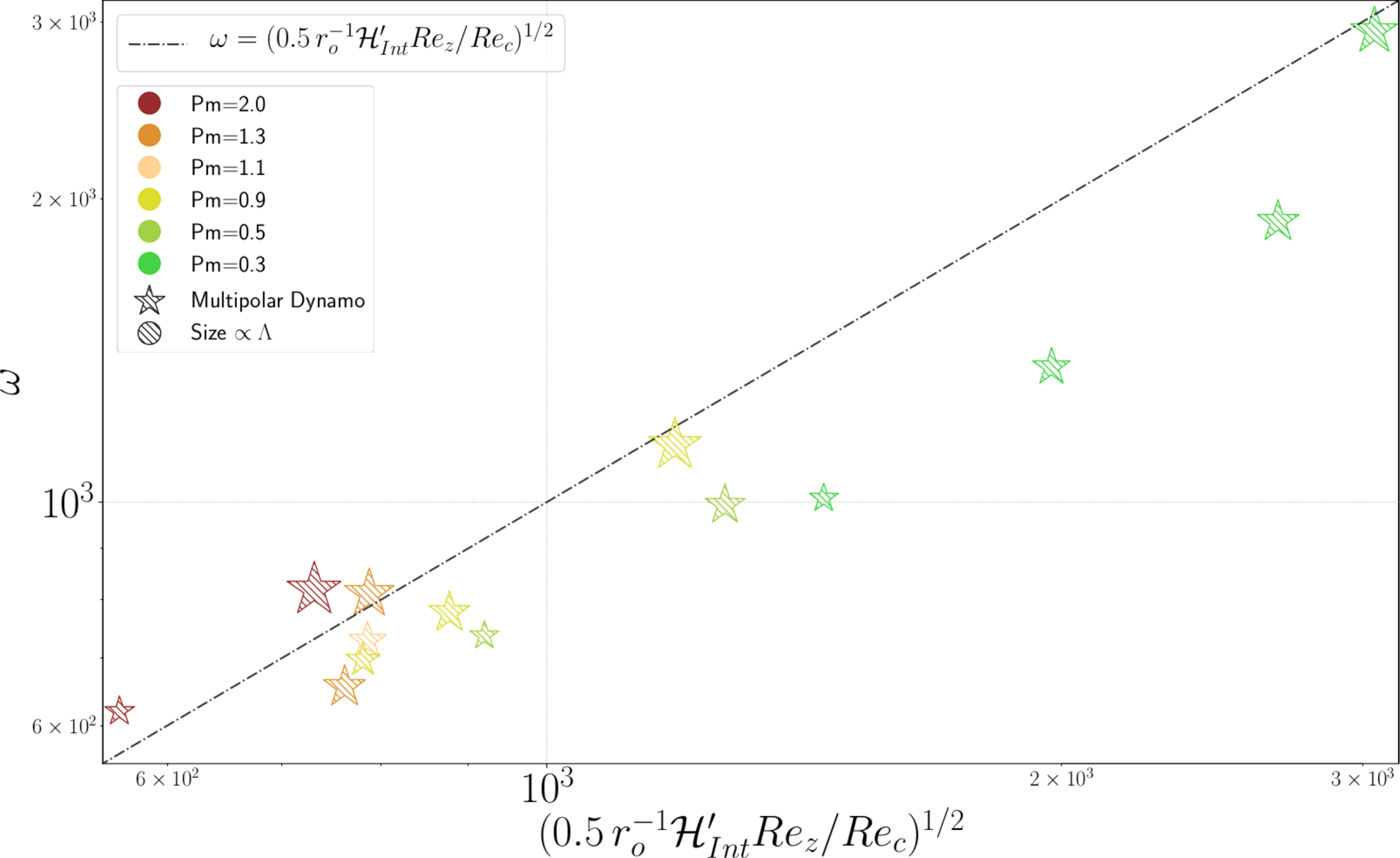}}
	\caption{
	Measured frequencies of oscillatory dynamos as a function of the frequency predicted by the Parker dynamo wave dispersion relation (\ref{eq:Parker_wave_number}), for all of our dynamos.
    Different colors correspond to different magnetic Prandtl numbers and the dashed line corresponds to the expected relation~(\ref{eq:Parker_wave_number}).
	}
	\label{fig:Summary-Laws_Omega-vs-Rezon}
\end{figure*}

Figure~\ref{fig:Summary-Laws_Omega-vs-Rezon} presents the resulting comparison between the oscillation periods extracted from our dynamos and those predicted for the Parker dynamo wave frequencies (above expression).
We find an overall agreement similar to what has already been found in 3D dynamo studies \citep{busse2006parameter,schrinner2011oscillatory,gastine2012dipolar} even though some of the individual retrieved frequencies can be offset by up to a factor two compared to the theory.
This is not unreasonable given all the approximations underlying the derivation of Eq.~(\ref{eq:Parker_wave_number}) so we can conclude that the observed oscillations in our dynamos can indeed be attributed to Parker dynamo waves.

\subsection{Comparison with 3D dynamos}
\label{sec:comparisont_3D-QG}

In order to compare our results with more conventional 3D simulations, we have computed a series of cases varying $Pm$ from $Pm=0.5$ to $Pm=0.05$ at $Ek=3 \times 10^{-5}$, $Pr=10^{-1}$, $Ra = 2 \times 10^{7} \sim 20\,Ra_c$ using a full 3D method \citep[{\tt MagIC},][]{wicht2002inner}.
Here we compare these results with a hybrid QG-3D case at $Ek=3 \times 10^{-5}$, $Pm=0.5$, $Pr=10^{-1}$, $Ra = 2 \times 10^{7} \sim 20\,Ra_c$.

\begin{figure*}
\centering{
	\includegraphics[width=0.49\linewidth]{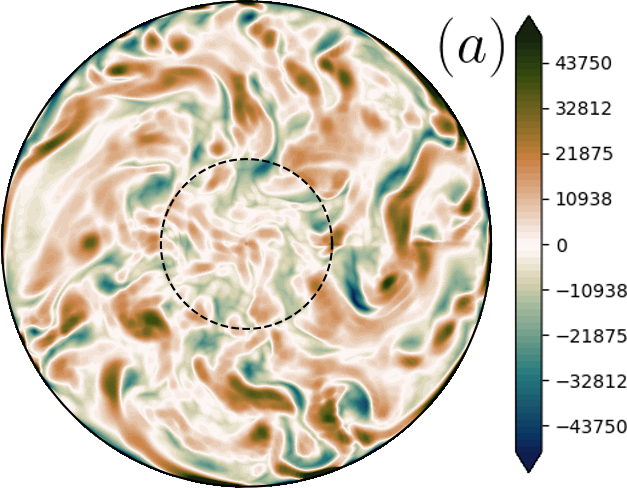}
	\includegraphics[width=0.49\linewidth]{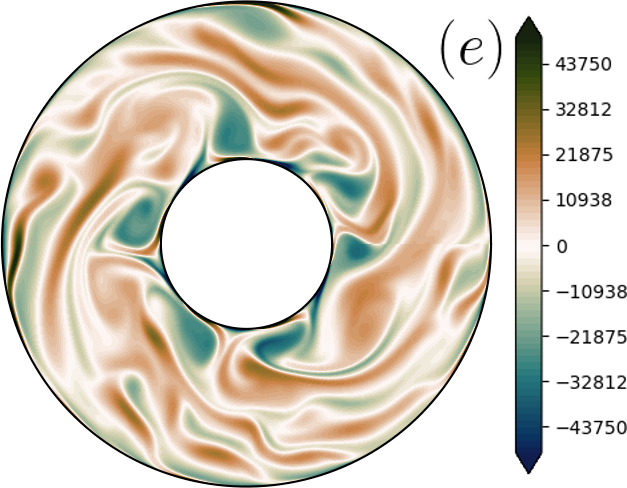}}
\centering{
	\includegraphics[width=0.24\linewidth]{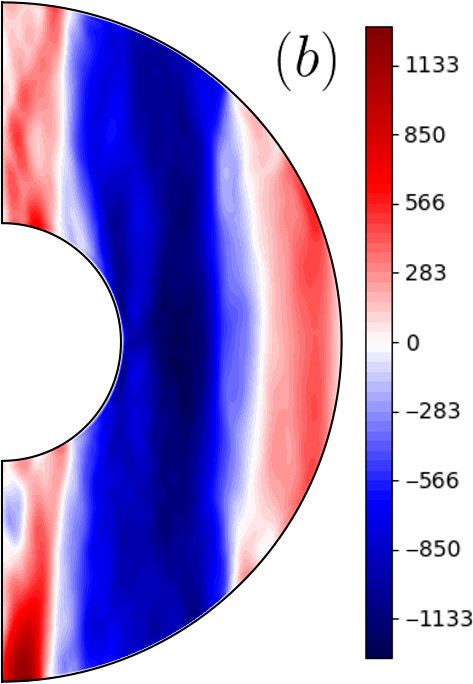}
	\includegraphics[width=0.24\linewidth]{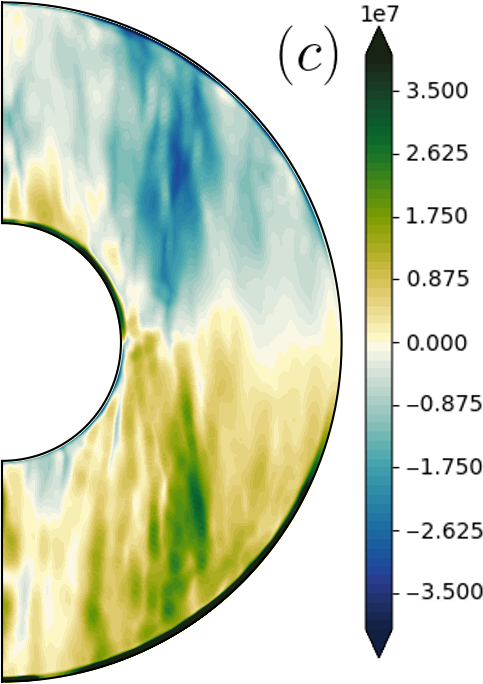}
	\includegraphics[width=0.24\linewidth]{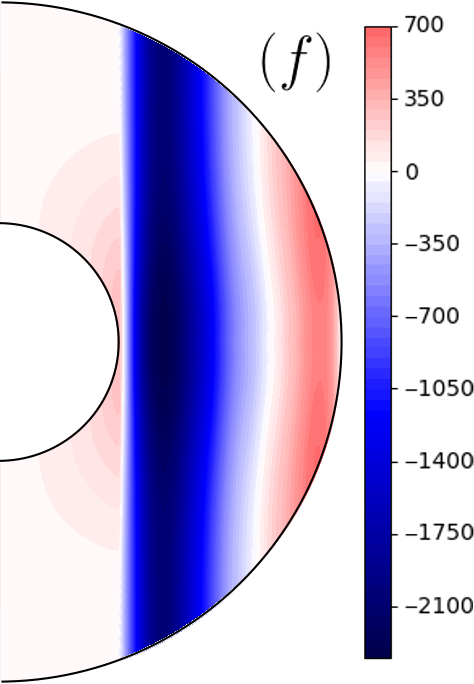}
	\includegraphics[width=0.24\linewidth]{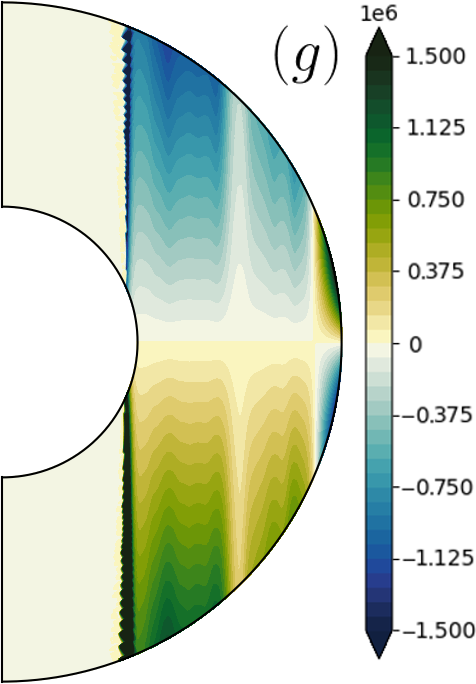}}
\centering{
	\includegraphics[width=0.49\linewidth]{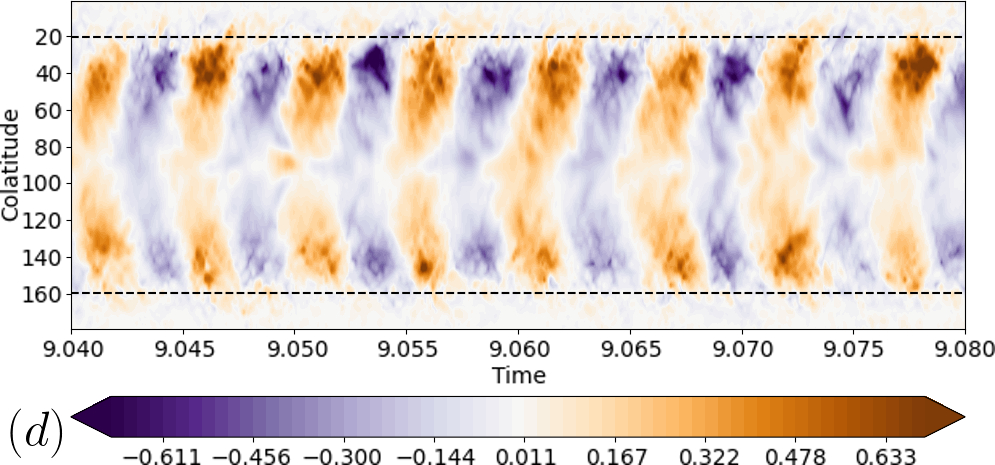}
	\includegraphics[width=0.49\linewidth]{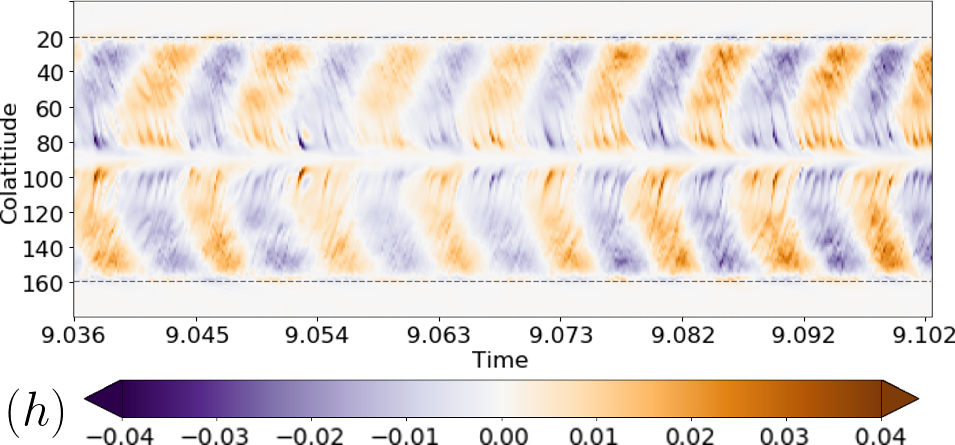}}
	\caption{
     Comparison of results from a 3D dynamo simulation computed at $Ek=3 \times 10^{-5}$, $Pr=0.1$, $Pm=0.1$, $Ra=20 \times Ra_c$ (panels a to d) and a hybrid QG-3D case at $Ek=3 \times 10^{-5}$, $Pr=0.1$, $Pm=0.5$, $Ra=20 \times Ra_c$ (panels e to h).
     (a) and (e) show snapshot of the $z$-averaged vorticity,
     (b) and (f) meridional section of the $\phi$-averaged azimuthal 3D velocity,
     (c) and (g) meridional section of the  $\phi$-averaged 3D helicity,
     (d) and (h) time series of the axisymmetric azimuthal magnetic field at $r\sim 0.92\,r_o$.
     }
	\label{fig:Pm0o1-0o5_Ra20_Comp-3D-QG}
\end{figure*}

Figure~\ref{fig:Pm0o1-0o5_Ra20_Comp-3D-QG} displays results for one of these 3D simulations, computed at $Pm=0.1$ (left column) and for the hybrid QG-3D simulation computed at $Pm=0.5$ (right column).
Snapshots of the $z$-averaged vorticity (panels a and e), meridional sections of the $\phi$-averaged azimuthal 3D velocity (panels b and f), meridional sections of the  $\phi$-averaged kinetic helicity $\overline{{\bm u}_\text{3D} \cdot \nabla \times {\bm u}_\text{3D}}$ (panels c and g), and time series of the axisymmetric azimuthal magnetic field at $r\sim 0.92\,r_o$ (panels d and h) are presented for both the 3D and hybrid cases.

We stress that the comparison involves different $Pm$ for both setups.
This is necessary because the onset for dynamo action happens at a lower $Pm$ with the 3D method \citep[see {\it e.g.},][]{petitdemange2018systematic}.
Hence, the chosen parameters compare results at a similar level of supercriticality with respect to the dynamo onset.
If one instead considers 3D and hybrid QG-3D results at the same control parameters, qualitative differences are found -- the solution is already bi-stable in the 3D case with a strong dipolar field, while a multipolar solution is only found when starting from a very small seed magnetic field.

Despite the difference in $Pm$, qualitatively similar solutions are then obtained when we are close to the onset of dynamo action for both configurations.
Figure~\ref{fig:Pm0o1-0o5_Ra20_Comp-3D-QG}~(a-e) show convective patterns that are close to a non-magnetic case with large length-scale and azimuthally elongated $z$-averaged vorticity patterns.
Both the 3D and hybrid QG-3D cases belong to the weak-field multipolar-branch dynamo with the kinetic energy that is greater than the magnetic energy, and the toroidal magnetic field that is much larger than the poloidal magnetic field.

Figure~\ref{fig:Pm0o1-0o5_Ra20_Comp-3D-QG}~(b-f) shows that the zonal flow patterns are similar in the two simulations, with a comparable level of geostrophy and similar radial gradients outside of the tangent cylinder, although the zonal flow is slightly stronger in the hybrid QG-3D case.
The largest differences between the two cases can be observed in their respective helicities (Fig.~\ref{fig:Pm0o1-0o5_Ra20_Comp-3D-QG}~c-g), with a maximum helicity that is about one order of magnitude lower in the hybrid QG-3D case and a change of sign towards low latitudes at the outer boundary that is not visible in the 3D case.
The spatial distribution of the helicity nevertheless remains comparable in the two cases with a segregation between mostly negative helicity in the northern hemisphere and mostly positive helicity in the southern hemisphere, similar to results previously reported for 3D simulations \citep[see, for example][]{davidson_ranjan_2018}.

Finally, we can see in Fig.~\ref{fig:Pm0o1-0o5_Ra20_Comp-3D-QG}~(d-h) that both the 3D and hybrid QG-3D simulations display similar Parker-wave oscillations although the magnetic field is equatorially symmetric rather than antisymmetric and contains more small scale features in the 3D case.
There is also much more magnetic energy in the 3D case with $\text{max} \overline{B_\phi^\text{3D}} > 10 \times \text{max} \overline{B_\phi^{\text{Hyb}}}$.
A rough estimation of the Parker cycles gives an oscillation period of $1$ cycle per $5 \times 10^{-3} \tau_\nu$ for the 3D case and $1$ cycle per $8.3 \times10^{-3} \tau_\nu$ for the hybrid QG-3D case, consistent with the fact that $Re_c$ and $Re_z$ are similar in the two simulations but ${\cal H}_\text{3D} \gg {\cal H}_\text{Hyb}$ (see Eq.~\ref{eq:Parker_wave_number}).

At $Pm=0.5$, $Ra \sim 20\,Ra_c$, we have computed the total r.m.s. kinetic helicity -- as a proxy of the amplitude of the $\alpha$-effect -- for our weak field hybrid QG-3D case, as well as for a multipolar weak field 3D case, and a dipolar strong field 3D case (at these parameters the 3D run is bi-stable).
Both 3D dynamos feature an average rms helicity roughly one order of magnitude larger than the hybrid QG-3D case (not shown).
There is clearly a global lack of helicity in the hybrid QG-3D case compared with the 3D case, even if the spatial segregation of helicity remains as expected.

Overall, we conclude that although the comparison of 3D and hybrid QG-3D dynamos is not straightforward, because they give different solutions when run at the same control parameters, the hybrid QG-3D and the 3D methods do produce qualitatively similar results when considered at the same distance from the onset of dynamo action, and provided the 3D case is started from a weak seed.
The major difference is that the hybrid QG-3D model involves much lower levels of kinetic heliticy.

\section{Discussion and conclusions}
\label{sec:Conclusion}

The results above demonstrate that it is possible to obtain dynamo action in a hybrid QG-3D magnetohydrodynamic model of rapidly-rotating convection in a spherical shell geometry, despite the strong assumptions described in section \ref{sec:Method} concerning the velocity field.
We have found several self-sustained multipolar dynamos in the parameter range $Ek = 3 \times 10^{-5}$, $Pr=0.1 - 1$, $Ra \sim (5 - 100)\,Ra_c$ and $Pm = 0.1- 2.0$ and have performed a detailed benchmarking of the Lorentz force in our setup as described in Appendix \ref{sec:validation}.

Focusing on simulations conducted at $Ek = 3 \times 10^{-5}$, $Pr=0.1$ and a range of $Pm$ and $Ra/Ra_c$, we found QG dynamos characterised by a low magnetic to kinetic energy ratio ${\cal M}$, a multipolar magnetic field, dominated by a toroidal field that is produced by an $\Omega$-effect sustained by strong zonal winds and with only a relatively weak poloidal field.
We have presented evidence for time-dependence in the form of dynamo waves in these solutions, similar to 3D models where the zonal flow plays an important role \citep[{\it e.g.},][]{schrinner2011oscillatory,simitev2012far,dietrich2013hemispherical,sheyko2016magnetic}.
A similar weak multipolar dynamo branch has been found in 3D models in a variety of set-ups \citep[{\it e.g.},][]{christensen2006scaling,schrinner2012dipole} and it seems that it is difficult to have both a strong dipolar magnetic field and a strong zonal flow \citep{gastine2012dipolar,duarte2013anelastic}.
Although, we have not conducted a thorough analysis of the parameter space at $Pr=1$, for the cases tested we found similar results with weak and multipolar magnetic field solutions, suggesting that having $Pr=0.1$ rather than 1 is not crucial to the form of the dynamos reported here.
A more detailed study focused on systematically varying the Prandtl number would be needed to fully quantify its role in the dynamo mechanism.

An initial motivation of this study was to test whether Earth-like dynamos could be achieved within a hybrid QG-3D convection-driven dynamo setup.
We have found no examples of dipole-dominated dynamos.
In contrast to 3D dynamos the poloidal field in our dynamos always remains weak compared with the toroidal field.
This remains true, even at other values of $Pr$ ({\it e.g.} runs at $Pr=1$, not shown), which suggests that our model may lack some important ingredient in the field generation cycle that operates in dipole-dominated 3D dynamos.
\citet{schaeffer2016can} previously found it was necessary to add an extra source of induction (magnetic pumping) in order to produce kinematic dynamos from observation-based QG flows.
We suspect that a lack of $\alpha$-effect associated with the lower level of helicity found in our dynamos compared to the 3D method is the main reason for their observed higher dynamo thresholds and possibly for the absence of a dipole-dominated branch in our configuration.
On the contrary, we expect that if the zonal component of the flow was removed or damped, we would simply lose the dynamo action because of the loss of $\Omega$-effect due to the strong zonal flow in the hybrid QG-3D set-up.
The absence of equatorially antisymmetric axial flows and the associated missing correlations between $u_z$ and the temperature $T_\text{3D}$ have been shown to play a rather significant role in a lack of convective power already observed in the non-magnetic configuration investigated in \citep{barrois2022comparison}.
These components are likely to also play a role in the helicity production \citep{dranjan_2020_segregation_helicity}.

A number of avenues can be envisaged for extending our model in order to enhance poloidal field generation.
One obvious option is to add a simple $\alpha$-effect term in the induction equation \citep[{\it e.g.},][]{chan2001non}, but such simple functional forms are difficult to justify in terms of the underlying convection.
Another option would be to follow \citet{schaeffer2016can} and implement a magnetic pumping whereby the velocity field is modified such that its helicity is enhanced based on an assumed toroidal magnetic field geometry \citep{sreenivasan_jones_2011,sreenivasan2018scale}.
Slow MAC or MC waves might also be an important source of helicity for producing a dipolar dynamo \citep{varma2022role}, but is unclear at the moment how best to parameterize their effect on the helicity, especially in the regions where the magnetic field is strong and heterogeneous.
A final possibility could be to include a simple form of $\alpha$-effect associated with helical waves propagating in the axial direction away from the equatorial plane where they are forced by turbulent convection.
Davidson and co-workers have explored the hypothesis that such waves, forced by convection, can play a role in generating dipolar magnetic fields \citep[for an overview see][]{davidson_ranjan_2018}.
The axial averaging applied in our setup removes the dynamo action associated with such helical waves; \citet{davidson_ranjan_2015_helicalwaves2} set out how such an $\alpha$-effect can be parameterized based on the kinetic energy of the flow in the equatorial plane.

An advantage of our hybrid QG-3D approach for the low magnetic Prandtl number regime of planetary cores is that it can treat the small scale velocity field efficiently within a QG framework while retaining a correct description of the 3D magnetic field and its boundary conditions.
However further numerical work is needed before our model can be applied to this regime.
So far, all our dynamos involved relatively weak Lorentz force and the energy of the magnetic field is much less than that associated with the velocity field.
Moreover, in the present implementation, because of challenges associated with the tangent cylinder discontinuity and because of the rather crude interpolation schemes used to move between the QG and the 3D grids, a very large number of points was needed for accurate and stable computations.
To take better advantage of the hybrid QG-3D approach with very different 3D and QG grid sizes it may be necessary to more carefully account for the action of the large length-scale magnetic field on the small length-scale velocity field, for example along the lines suggested by \citet{schaeffer2006quasi}.

Returning to the geophysical context, we conclude that our hybrid QG-3D model seems incapable of producing Earth-like (strong-field, dipole-dominated) dynamos.
This suggests, in agreement with the earlier findings of \citet{schaeffer2016can}, that something important for geodynamo action is lost in moving between 3D flows and the simplified QG flows considered here.
If hybrid QG-3D models are to be used to study the long-term behaviour of the geodynamo it will be necessary to find a principled scheme for parameterizing these missing effects, which may be related to structures in the axial flow component and their helicity.
Hybrid QG-3D models could however already prove to be a valuable tool for studying the short-term behaviour of the geodynamo, on timescales shorter than the convective timescales when the dynamics is dominated by QG hydromagnetic waves \citep{aubert2018,aubert2022taxonomy,gillet_2022}.
On these timescales, the dynamo-generated field can be considered steady and could be imposed, for example, based on results from a 3D simulation producing an Earth-like field \citep[{\it e.g.},][]{aubert2023state}.
The hybrid QG-3D model is capable of efficiently representing both QG wave flows and related 3D magnetic field perturbations and has the potential to be significantly faster than full 3D simulations for studying such waves.

\section*{Acknowledgements}

We thank two anonymous reviewers and the editor Richard Holme for their constructive comments which have helped to improve the manuscript.
Nathana\"{e}l Schaeffer and C\'{e}line Guervilly are thanked for helpful discussions and for sharing their code for the $z$-averaging functions and for the $z$-integration of the thermal wind.
QG and hybrid QG-3D numerical computations using the \texttt{pizza} code were performed at DTU Space, using the Humboldt and Larmor CPU clusters.
OB and CCF were supported by the European Research Council (ERC) under the European Union’s Horizon 2020 research and innovation program (grant agreement No. 772561).

\section*{Data availability}

All our data are available upon reasonable request to the corresponding author and some key parameters from our whole data-set are already included in the present article.
The codes used (\texttt{pizza}) is freely available (at \url{http://www.github.com/magic-sph/pizza/tree/hybrid_QG-3D}) under the GNU GPL v3 license.

\bibliography{artbib}

\bibliographystyle{gji}

\onecolumn

\appendix
\section{Results of numerical simulations}
\label{sec:Append-A-Results}

\begin{longtable}{p{0.11\textwidth} p{0.045\textwidth} p{0.035\textwidth} p{0.04\textwidth} p{0.065\textwidth} p{0.125\textwidth} p{0.125\textwidth} p{0.035\textwidth} p{0.065\textwidth} p{0.19\textwidth}}
\caption{Summary of the hybrid QG-3D numerical simulations computed in this study at $Ek = 3 \times 10^{-5}$, $Pr = 0.1$, using $\eta = r_i/r_o = 0.35$.
$Ra$ is the Rayleigh number (the supercriticality $Sc = Ra / Ra_c$ is also provided, where $Ra_c$ is the thermal convection critical value), $Pm$ is the magnetic Prandtl number, $Nu$ is the Nusselt number, $Rm$ is the magnetic Reynolds number, $\Lambda$ is the Elsasser number, ${\cal M} =  \widehat{E}_\text{mag} / \widehat{E}_\text{kin}$ is the magnetic to kinetic energy ratio, $f_\text{tor} = \widehat{E}_\text{tor}/\widehat{E}_\text{mag}$ is the toroidal to the total magnetic field energy ratio, $\ell_H$ and $m_H$ are the cut-off degree and azimuthal wavenumber below which the hyperdiffusion has no effect, and $(N_s, N_m)/(N_r, \ell_\text{max})$ are the grid-sizes of the run.
The simulations with a $^\bigstar$ symbol in the second column are the growing dynamos we found.}
\label{tab:run_list} \\
\hline
$Ra$ & $Sc$ & $Pm$ & $Nu$ & $Rm$ & $\Lambda$ & $\mathcal{M}$ & $f_\text{tor}$ & $m_H/\ell_H$ & $(N_s, N_m)/(N_r, \ell_\text{max})$ \\
\hline
\endfirsthead

\hline 
$Ra$ & $Sc$ & $Pm$ & $Nu$ & $Rm$ & $\Lambda$ & $\mathcal{M}$ & $f_\text{tor}$ & $m_H/\ell_H$ & $(N_s, N_m)/(N_r, \ell_\text{max})$ \\
\hline
\endhead

\hline
\multicolumn{9}{c}{Continued on next page $\ldots$} \\
\hline
\endfoot

\hline
\hline
\endlastfoot

$5.00 \times 10^6$  & $4.9$ & $0.9$ & $1.19$ & $334.0$ & $2.08 \times 10^{-6}$ & $4.86 \times 10^{-7}$ & $0.81$ & $-/-$ & $(193, 192)/(145, 144)$ \\
$5.00 \times 10^6$  & $4.9$ & $1.3$ & $1.19$ & $482.9$ & $8.61 \times 10^{-9}$ & $1.39 \times 10^{-9}$ & $0.79$ & $144/-$ & $(193, 192)/(145, 144)$ \\
$5.00 \times 10^6$ & $4.9$ & $2.0$ & $1.19$ & $740.4$ & $1.66 \times 10^{-6}$ & $1.76 \times 10^{-7}$ & $0.82$ & $144/-$ & $(385, 768)/(149, 148)$ \\
$1.00 \times 10^7$ & $9.7$ & $0.5$ & $1.32$ & $374.7$ & $1.09 \times 10^{-6}$ & $1.13 \times 10^{-7}$ & $0.77$ & $144/-$ & $(385, 768)/(149, 148)$ \\
$1.00 \times 10^7$ & $9.7$ & $0.9$ & $1.32$ & $670.9$ & $5.41 \times 10^{-6}$ & $3.15 \times 10^{-7}$ & $0.89$ & $144/-$ & $(385, 768)/(149, 148)$ \\
$1.00 \times 10^7$ & $9.7$ & $1.3$ & $1.32$ & $976.1$ & $8.97 \times 10^{-3}$ & $3.56 \times 10^{-4}$ & $0.97$ & $144/-$ & $(385, 768)/(149, 148)$ \\
$1.00 \times 10^7$  & $9.7^\bigstar$ & $2.0$ & $1.32$ & $1456.4$ & $2.48 \times 10^{-1}$ & $6.79 \times 10^{-3}$ & $0.98$ & $-/-$ & $(145, 144)/(145, 144)$ \\
$1.65 \times 10^7$ & $16.0^\bigstar$ & $0.9$ & $1.40$ & $894.3$ & $1.31 \times 10^0$ & $4.28 \times 10^{-2}$ & $0.96$ & $-/-$ & $(129, 192)/(129, 192)$ \\
$1.66 \times 10^7$ & $16.1$ & $0.3$ & $1.42$ & $347.0$ & $4.80 \times 10^{-7}$ & $3.48 \times 10^{-8}$ & $0.78$ & $144/-$ & $(385, 768)/(149, 148)$ \\
$1.66 \times 10^7$ & $16.1$ & $0.5$ & $1.42$ & $578.2$ & $7.31 \times 10^{-5}$ & $3.18 \times 10^{-6}$ & $0.95$ & $144/-$ & $(385, 768)/(149, 148)$ \\
$1.66 \times 10^7$ & $16.1^\bigstar$ & $0.9$ & $1.41$ & $909.1$ & $1.27 \times 10^0$ & $4.03 \times 10^{-2}$ & $0.96$ & $144/-$ & $(385, 768)/(149, 148)$ \\
$1.66 \times 10^7$ & $16.1^\bigstar$ & $1.1$ & $1.41$ & $1029.7$ & $2.51 \times 10^0$ & $7.55 \times 10^{-2}$ & $0.96$ & $-/-$ & $(145, 192)/(145, 192)$ \\
$1.66 \times 10^7$ & $16.1^\bigstar$ & $1.3$ & $1.41$ & $1174.0$ & $3.68 \times 10^0$ & $1.00 \times 10^{-1}$ & $0.96$ & $-/-$ & $(145, 192)/(145, 192)$ \\
$1.66 \times 10^7$ & $16.1^\bigstar$ & $2.0$ & $1.40$ & $1747.4$ & $8.02 \times 10^0$ & $1.51 \times 10^{-1}$ & $0.95$ & $288 /-$ & $(289, 416)/(193, 288)$ \\
$2.00 \times 10^7$ & $19.4$ & $0.1$ & $1.46$ & $135.6$ & $3.58 \times 10^{-9}$ & $5.67 \times 10^{-10}$ & $0.69$ & $144/-$ & $(385, 768)/(149, 148)$ \\
$2.00 \times 10^7$ & $19.4$ & $0.3$ & $1.46$ & $407.1$ & $8.20 \times 10^{-6}$ & $4.32 \times 10^{-7}$ & $0.89$ & $144/-$ & $(385, 768)/(149, 148)$ \\
$2.00 \times 10^7$ & $19.4^\bigstar$ & $0.5$ & $1.45$ & $665.8$ & $7.15 \times 10^{-2}$ & $2.35 \times 10^{-3}$ & $0.95$ & $-/-$ & $(97, 128)/(97, 128)$ \\
$2.00 \times 10^7$ & $19.4^\bigstar$ & $0.9$ & $1.44$ & $913.0$ & $3.46 \times 10^0$ & $1.08 \times 10^{-1}$ & $0.95$ & $144/-$ & $(145, 180)/(145, 144)$ \\
$2.00 \times 10^7$ & $19.4^\bigstar$ & $1.3$ & $1.45$ & $1274.1$ & $6.14 \times 10^0$ & $1.42 \times 10^{-1}$ & $0.96$ & $-/-$ & $(145, 192)/(145, 192)$ \\
$3.00 \times 10^7$ & $29.1$ & $0.1$ & $1.55$ & $188.9$ & $3.86 \times 10^{-7}$ & $3.15 \times 10^{-8}$ & $0.73$ & $144/-$ & $(385, 768)/(149, 148)$ \\
$2.99 \times 10^7$ & $29.0$ & $0.2$ & $1.54$ & $375.3$ & $2.51 \times 10^{-7}$ & $1.04 \times 10^{-8}$ & $0.77$ & $-/-$ & $(385, 960)/(385, 320)$ \\
$3.00 \times 10^7$ & $29.1^\bigstar$ & $0.3$ & $1.54$ & $557.8$ & $9.75 \times 10^{-2}$ & $2.74 \times 10^{-3}$ & $0.93$ & $-/-$ & $(97, 128/(97, 128)$ \\
$3.00 \times 10^7$ & $29.1^\bigstar$ & $0.5$ & $1.52$ & $691.2$ & $3.16 \times 10^0$ & $9.58 \times 10^{-2}$ & $0.94$ & $256/-$ & $(433, 864)/(193, 256)$ \\
$3.00 \times 10^7$ & $29.1^\bigstar$ & $0.9$ & $1.57$ & $1283.8$ & $7.61 \times 10^0$ & $1.21 \times 10^{-1}$ & $0.92$ & $224/224$ & $(433, 480)/(193, 256)$ \\
$5.00 \times 10^7$ & $48.5$ & $0.1$ & $1.66$ & $275.7$ & $3.84 \times 10^{-7}$ & $1.47 \times 10^{-8}$ & $0.79$ & $-/-$ & $(433, 800)/(149, 148)$ \\
$5.00 \times 10^7$ & $48.5$ & $0.2$ & $1.66$ & $550.9$ & $6.91 \times 10^{-6}$ & $1.33 \times 10^{-7}$ & $0.84$ & $816/-$ & $(433, 864)/(149, 148)$ \\
$4.99 \times 10^7$ & $48.4^\bigstar$ & $0.3$ & $1.62$ & $640.5$ & $2.31 \times 10^0$ & $4.91 \times 10^{-2}$ & $0.91$ & $576/144$ & $(481, 912)/(193, 256)$ \\
$8.00 \times 10^7$ & $77.7$ & $0.1$ & $1.73$ & $376.5$ & $7.80 \times 10^{-7}$ & $1.60 \times 10^{-8}$ & $0.83$ & $144/-$ & $(577, 864)/(193, 192)$ \\
$8.00 \times 10^7$ & $77.7$ & $0.2$ & $1.73$ & $754.4$ & $1.07 \times 10^{-4}$ & $1.10 \times 10^{-6}$ & $0.86$ & $144/-$ & $(577, 864)/(193, 192)$ \\
$8.00 \times 10^7$ & $77.7^\bigstar$ & $0.3$ & $1.66$ & $836.0$ & $3.65 \times 10^{0}$ & $4.55 \times 10^{-2}$ & $0.92$ & $192/-$ & $(289, 384)/(193, 192)$ \\
$1.03 \times 10^8$ & $100$ & $0.1$ & $1.76$ & $439.9$ & $4.42 \times 10^{-5}$ & $6.65 \times 10^{-7}$ & $0.84$ & $768/-$ & $(577, 864)/(193, 192)$ \\
$1.03 \times 10^8$ & $100$ & $0.2$ & $1.74$ & $861.7$ & $9.89 \times 10^{-3}$ & $1.94 \times 10^{-4}$ & $0.94$ & $768/-$ & $(577, 912)/(193, 256)$ \\
$1.03 \times 10^8$ & $100^\bigstar$ & $0.3$ & $1.69$ & $982.1$ & $5.44 \times 10^{0}$ & $4.92 \times 10^{-2}$ & $0.94$ & $192/-$ & $(289, 384)/(193, 192)$ \\
\end{longtable}

\section{Code Validation and benchmarks}
\label{sec:validation}

\subsection{Benchmark of the Lorentz force computation}
\label{sec:LF-bench}

The determination of Eq.~(\ref{eq:LF-QG_3D}) relies on the computation of $({\bf j} \times {\bf B})$ which depends on ${\bm \nabla} \times {\bf B}$ and ${\bf B}$.
These quantities are computed using \texttt{SHTns}\footnote{\url{https://bitbucket.org/nschaeff/shtns}} functions, already validated in \citep{schaeffer2013efficient} and widely used in many codes.
Given any fields $(B_\text{pol}, B_\text{tor})$ in spectral space, and their radial-derivatives, the \texttt{SHTns} routines directly provide $(B_r, B_\theta, B_\phi)$ and $(j_r, j_\theta, j_\phi)$ on the 3D-physical-grid.
The remaining part of the process involves the computation of the non-linear products $({\bf j} \times {\bf B})$, and the $z$-averaging of $\partial_s \left[ s ({\bf j} \times {\bf B})_\phi \right]$, $({\bf j} \times {\bf B})_\phi$ and $({\bf j} \times {\bf B})_s$ which are computed on the physical grid.
These quantities are then sent to the spectral space using {\tt fft} functions where the $\phi$-derivative is performed before assembling all the terms to obtain Eq.~(\ref{eq:LF-QG_3D}).

\subsubsection{Summary of the steps involved in the Lorentz force computation}

As a summary, the steps involved in our computation of the QG-Lorentz force are presented in the following steps
\begin{align}
\label{eq:LF-steps-by-steps}
\text{\bf{physical grid}} \nonumber \\
\bf{1)}\; (B_r, B_\theta, B_\phi) \times (j_r, j_\theta, j_\phi) \rightarrow ({\bf j} \times {\bf B}) \cdot ({\bm e}_r, {\bm e}_\theta, {\bm e}_\phi)\,; \; \text{non-linear products}\,, \nonumber \\
\bf{2)}\; ({\bf j} \times {\bf B})_r, ({\bf j} \times {\bf B})_\theta \rightarrow ({\bf j} \times {\bf B})_s\,; \; \text{linear operation}\,, \nonumber \\
\bf{3)}\; ({\bf j} \times {\bf B})_\phi \rightarrow \dfrac{\partial }{\partial s} \left[ s ({\bf j} \times {\bf B})_\phi \right]\,; \; s\text{-derivative}\,, \nonumber \\
\bf{4)}\; \left\lbrace\begin{aligned}
({\bf j} \times {\bf B})_\phi \rightarrow \left< ({\bf j} \times {\bf B})_\phi \right> \\
({\bf j} \times {\bf B})_s \rightarrow \left< ({\bf j} \times {\bf B})_s \right> \\
\dfrac{\partial }{\partial s} \left[ s ({\bf j} \times {\bf B})_\phi \right] \rightarrow \left< \partial_s \left[ s ({\bf j} \times {\bf B})_\phi \right] \right> \\
\end{aligned}\right.\,; \; z\text{-averages}\,, \nonumber \\
\text{\bf{physical-to-spectral space}} \nonumber \\
\bf{5)}\; \left\lbrace\begin{aligned}
\left< ({\bf j} \times {\bf B})_\phi \right>  \rightarrow \widetilde{\left< ({\bf j} \times {\bf B})_\phi \right>} \\
\left< ({\bf j} \times {\bf B})_s \right>  \rightarrow \widetilde{\left< ({\bf j} \times {\bf B})_s \right>} \\
\left< \partial_s \left[ s ({\bf j} \times {\bf B})_\phi \right] \right> \rightarrow \widetilde{\left< \partial_s \left[ s ({\bf j} \times {\bf B})_\phi \right] \right>} \\
\end{aligned}\right.\,; \; \text{Fourier transforms}\,, \nonumber \\
\text{\bf{spectral space}} \nonumber \\
\bf{6)}\; \widetilde{\left< ({\bf j} \times {\bf B})_s \right>} \rightarrow \it i\,m \widetilde{\left< ({\bf j} \times {\bf B})_s \right>}\,; \; \phi\text{-derivative}\,, \nonumber \\
\bf{7)}\; \left\lbrace\begin{aligned}
\widetilde{\left< ({\bf j} \times {\bf B})_\phi \right>} \rightarrow \dfrac{1}{Ek\,Pm}\widetilde{\left< ({\bf j} \times {\bf B})_\phi \right>}_{m=0} = \widetilde{{\bf F}_{{\cal L}\,,\,\overline{u_\phi}}}\,; \; \text{assemble Lorentz force for the zonal-flow Eq.}\,, \\
(\widetilde{\left< \partial_s \left[ s ({\bf j} \times {\bf B})_\phi \right] \right>}, i\,m\widetilde{\left< ({\bf j} \times {\bf B})_s \right>}) \rightarrow \dfrac{1}{Ek\,Pm} \left( \dfrac{1}{s} \widetilde{\left< \partial_s \left[ s ({\bf j} \times {\bf B})_\phi \right] \right>} -\dfrac{i\,m}{s} \widetilde{\left< ({\bf j} \times {\bf B})_s \right>} \right) =  \widetilde{{\bf F}_{{\cal L}\,,\,\omega_z}} \; \\
\text{assemble Lorentz force for the vorticity Eq.}\,. \\
\end{aligned}\right.
\end{align}
Where the wide tildes $\widetilde{x}$ refers to the Fourier transform of any quantity $x$, that is the quantity $x$ in the spectral space.

\subsubsection{Analytical benchmark from an artificial ${\bf j} \times {\bf B}$ field}

As steps {\bf 1}) and {\bf 2}) in (\ref{eq:LF-steps-by-steps}) are respectively handled by \texttt{SHTns} routines and simply involve a linear product, we analytically validate our method from step {\bf 3}).
We need a field that is easily differentiable in $s$, easily integrated in $z$, periodic in $\phi$, cancels at the outer boundary, but is not trivial.
Following these constraints, we thus chose a completely artificial field, that reads
\begin{align}
\label{eq:jxB_analytical}
\left\lbrace\begin{aligned}
({\bf j} \times {\bf B})_\phi^\text{ref} (r, \theta, \phi) &= -\left(\dfrac{1}{2} + \cos\phi + \sin(4\phi)\right)\,\dfrac{\pi\,h^3}{s}\cos\left(\pi\,\dfrac{z}{2h}\right) \\
({\bf j} \times {\bf B})_s^\text{ref} (r, \theta, \phi) &= \sin^2\phi\,s(s_o-s)(s-s_i)\,z^2\\
({\bf j} \times {\bf B})_z^\text{ref} (r, \theta, \phi) &= \sin^2\phi\,s(s_o-s)(s-s_i)\,z
\end{aligned}\right.
\end{align}
leading to
\begin{align}
\label{eq:jxB-2-LF_analytical}
{\tt F}_{\cal L}^\text{ref} = \left\lbrace\begin{aligned}
&\left[1/2 + \cos\phi + \sin(4\phi)\right]\,8hs - \dfrac{2}{3}\cos\phi\sin\phi\,(s_o - s)(s - s_i)\,h^2 \\
&-\dfrac{h^3}{s}
\end{aligned}\right.\,.
\end{align}

\subsubsection{Relative error definition}
\label{sec:def_erel}

In order to discuss the validation and the accuracy of our numerical schemes, we define a relative error estimate, $e_\text{rel}$, as
\begin{align}
\label{eq:3D-analytic_error-rel}
e_\text{rel}(f) = \displaystyle\left[ \dfrac{ \left\lbrace (f_\text{ref} - f)^2 \right\rbrace_\text{QG} }{ \left\lbrace f_\text{ref}^2 \right\rbrace_\text{QG} } \right]^{1/2}\,,
\end{align}
where the brackets in the above equation correspond to an average over the annulus as in Eq.(\ref{eq:spatial_averages}).

\subsection{Results}

\begin{figure}
\centerline{
    \includegraphics[width=0.97\linewidth]{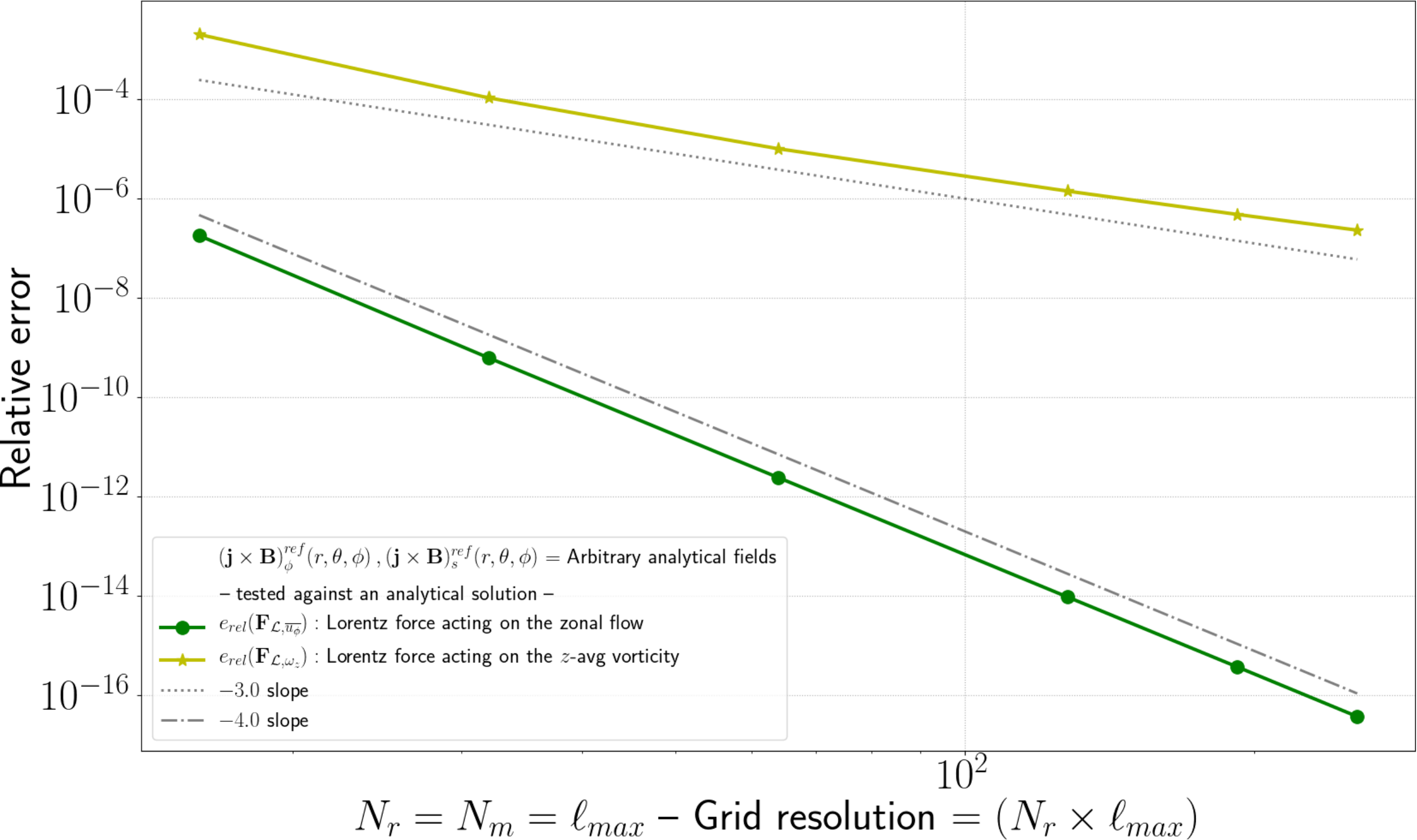}}
    \caption[]{
    Convergence of the relative error of the Lorentz force terms tested against an analytical solution.
    }
    \label{fig:comp-zavgVxjxBez_jxB_Analytical_errel}
\end{figure}

Figure~\ref{fig:comp-zavgVxjxBez_jxB_Analytical_errel} displays the convergence of the relative error for the Lorentz force terms as a function of the resolution.
We can see that the computation of the Lorentz force term acting on the zonal flow ${\bf F}_{{\cal L}\,,\,\overline{u_\phi}}$ (green curve) -- that only involves a $z$-averaging of $({\bf j} \times {\bf B})_\phi$ -- has an accuracy of order $4$.
Compared with \cite{barrois2022comparison}, for all the $z$-integration steps we have used in this work a Simpson rule of integration, with an order $4$ accuracy, and we satisfactorily retrieve the expected precision.
On the other hand, we find a global accuracy of order $3$ for the Lorentz force term acting on the vorticity ${\bf F}_{{\cal L}\,,\,\omega_z}$ (yellow curve) which involves additional operations (a $\phi$- and $s$-derivatives).

\subsection{Conclusion}

We have been able to retrieve the correct Lorentz force terms acting on both the $z$-averaged vorticity and the zonal-flow equations starting from an imposed artificial $({\bf j} \times {\bf B})$-field and we have additionally verified that these results were consistent with an independent method (not shown).

The overall accuracy of the computation of the Lorentz force is limited by a number of interpolating schemes -- {\it i.e.} a $s$-derivative scheme, a $z$-averaging scheme, and an {\tt ifft}.
We find an accuracy of order $4$ converging toward an average relative error of $10^{-16}$ (compared with an analytical solution) for the Lorentz force term acting on the zonal flow equation.
And we find a global accuracy of order $3$ converging toward an average relative error of $10^{-6}$ (compared with an analytical solution) for the Lorentz force term acting on the vorticity equation.
We thus consider the computation of Eq.~(\ref{eq:LF-QG_3D}) in our code to be validated.

\label{lastpage}

\end{document}